\documentclass[fleqn,10pt]{wlscirep}
\usepackage[utf8]{inputenc}
\usepackage[T1]{fontenc}

\usepackage{lineno}

\usepackage{booktabs}
\usepackage{multirow}
\usepackage{longtable}
\usepackage{multicol}
\usepackage{threeparttable}
\usepackage{tabularx}
\usepackage{array}

\newcolumntype{Y}{>{\raggedright\arraybackslash}X}

\makeatletter
\newcommand{\beginsupplementaryinformation}{%
  \clearpage
  \newgeometry{left=3cm,right=3cm,top=3cm,bottom=3cm}%
  \pagestyle{plain}%
  \setlength{\parindent}{15pt}%
  \setlength{\parskip}{0pt}%
  \renewcommand{\rmdefault}{ptm}%
  \normalfont
  \fontsize{11}{13.6}\selectfont
  \captionsetup{labelfont=bf,textfont=normalfont,labelsep=period,justification=justified,singlelinecheck=false}%
  \setcounter{figure}{0}%
  \setcounter{table}{0}%
  \renewcommand{\thefigure}{S\arabic{figure}}%
  \renewcommand{\thetable}{S\arabic{table}}%
  \setcounter{section}{0}%
  \setcounter{subsection}{0}%
  \setcounter{subsubsection}{0}%
  \titleformat{\section}{\normalfont\Large\bfseries}{\thesection}{0.75em}{##1}%
  \titleformat{name=\section,numberless}{\normalfont\Large\bfseries}{}{0em}{##1}%
  \titleformat{\subsection}{\normalfont\large\bfseries}{\thesubsection}{0.75em}{##1}%
  \titleformat{\subsubsection}{\normalfont\normalsize\itshape}{\thesubsubsection}{0.75em}{##1}%
  \titleformat{\paragraph}[runin]{\normalfont\normalsize\bfseries}{}{0pt}{##1}%
  \titlespacing*{\section}{0pt}{3.5ex plus 1ex minus .2ex}{2.3ex plus .2ex}%
  \titlespacing*{\subsection}{0pt}{3.25ex plus 1ex minus .2ex}{1.5ex plus .2ex}%
  \titlespacing*{\subsubsection}{0pt}{3.25ex plus 1ex minus .2ex}{1.5ex plus .2ex}%
  \titlespacing*{\paragraph}{0pt}{1ex}{1em}%
}

\newcommand{\supplementarytableofcontents}{%
  {\centering\large\bfseries Supplementary Notes\par}%
  \vspace{0.75\baselineskip}%
  \@starttoc{stc}%
}
\makeatother

\title{%
Planning for isolation? The role of urban form and function in shaping mobility in Brasília
}

\author[a,*]{Andrew Renninger}

\affil[a]{Centre for Advanced Spatial Analysis, University College London}

\affil[*]{Corresponding author: Andrew Renninger (E-mail: andrew.renninger.12@ucl.ac.uk)}

\begin{abstract}
Brasília offers a rare test of how urban form shapes experienced segregation. Built almost at once around modernist neighbourhood units, then expanded through planned satellites and informal peripheries, it lets us ask whether urban form turns mobility into mixing or into a more efficient engine of separation. We combine data on human mobility with urban morphometrics, amenities, road networks, along with enclosures and tessellations that capture segregation at the scales where access is structured: districts, neighbourhoods, blocks, and street-and-building cells. We find that segregation intensifies as resolution sharpens, from 0.282 at the district scale to 0.545 at the block scale, indicating that Brasília looks most integrated at coarse units and most segregated where everyday encounters are actually organised. Mobility softens home segregation for most users, but not symmetrically: poorer groups travel farther, while affluent groups remain the most selectively exposed. civic cores and mid-rise, mixed-use areas are the least segregated morphotypes, yet they occupy only a sliver of the metropolis. Elsewhere, rich lakefront suburbs and dense poor settlements reach similarly high segregation through opposite spatial logics. Amenities predict lower segregation, while barriers and enclosed residential interiors predict higher segregation. Built form explains more of this pattern than visit volume alone in the segregation models: integration is less a property of residential design than of shared destinations and porous connections. Planned capitals can build order without building isolation if they distribute mixing space rather than sequestering it.
\end{abstract}

\begin{document}

\flushbottom
\maketitle

%
\section*{Introduction}
In the 20th century, planners and architects developed theories and designs for the ideal city and attempted to implement them in ``new towns''---towns or even whole cities built on vacant land, free of existing streets, buildings and people who could block or constrain development \cite{peiser2021towns, hall2014cities}. Notable examples include Chandigarh in India, Canberra in Australia, and Milton Keynes in the United Kingdom. Yet while space for development was a primary concern \cite{howard1902garden, hall2014cities, hall1998sociable}, these new towns often packaged sets of assumptions about human behaviour and how people interact with the built and natural environments that comprise a city \cite{fishman1982utopias, lawhon2009determinism}. Many modernist plans pursued order not only through fast circulation, but through the design of the neighbourhood unit as the constituent element of urban life. The promise was access: a contained, walkable world in which houses, schools, parks, and key services were assembled into coherent districts, intended to produce community through proximity and routine encounter \cite{perry1929neighborhood, silver1985neighborhood}. At the metropolitan scale, these units were stitched together by a hierarchy of slower roads and faster corridors, reflecting a belief that cities could be made more efficient and governable by decomposing urban complexity into modular parts \cite{perry1929neighborhood,lawhon2009determinism}. Today, as Egypt \cite{reuters2024egypt,abusaada2023singularity} and Indonesia \cite{reuters2026nusantara,syaban2023capital} move administrative capitals away from crowded megacities, new towns are again playing an important role in urban thought and design. Many of these projects make similar assumptions and present familiar ideals, blending aesthetic visions of order with claims about human needs \cite{abusaada2022similarity, abusaada2023singularity, syaban2023capital, teo2020capital}. The neighbourhood unit that became the scalable and portable unit enabling many twentieth century plans has even returned as the ``15-minute city'' \cite{moreno2021minute, khavariangarmsir2023garden}. This makes planned cities worth revisiting not only as episodes in design history, but as unusually explicit theories of how urban life should be organised. 

While many studies have documented cases in which modernist town planning has transformed urban life---increasing dependence on the automobile \cite{cervero1997travel, ewing2010meta}, reducing pedestrian activity on streets and sidewalks \cite{appleyard1972streets, saelens2003walking}, and influencing both mental \cite{evans2003housing, guite2006wellbeing} and physical health \cite{frank2004obesity, ewing2003sprawl}---less attention has been paid to how these spatial logics shape the broader patterns of who interacts with whom in the city. The neighbourhood unit, for example, attempts to reduce the need to travel between communities for schooling, recreation and other services, and the multilane roads that link these units to downtowns allow drivers to avoid intervening communities en route to office districts. Recent work has begun to show that segregation is mediated not only by social composition but by urban form itself: block morphologies, land uses, neighbourhood conditions, activity-space types all shape opportunities for encounter and separation \cite{miranda2020shape, gao2023socio, sun2024social, kristensen2023urban, useche2024spatial}.

A growing body of literature documents ``experienced segregation''—our propensity to frequent locations that are in turn frequented by others like us, socio-economically \cite{moro2021mobility} or demographically \cite{athey2021estimating}—across diverse contexts \cite{liao2025socio}. Socio-demographic communities in a city will often form a ``city-within-a-city’’. In the purist display of these preferences, the wealthiest residents of London produce ``cloud spaces’’ or ``flowing enclaves’’—travelling between private clubs in private cars, for example—to avoid contact with the broader public \cite{atkinson2016limited}; the elite in London can spend as much as 80\% of the day in exclusive zones. Yet the phenomenon is not just practised by the elite and it is not unique to London: studies document similar dynamics between Catholics and Protestants in Northern Ireland \cite{davies2019networks, abdelmonem2015search} and Muslims and Jews in Israel \cite{schnell2014arab, rokem2018segregation}, even when commuting to the same place brings those groups into close proximity. Generally, people are more likely to go to venues popular within their own socio-economic class \cite{hilman2022socioeconomic}, but there tends to be a bias: when individuals do deviate from assortative mixing, they prefer points of interest with higher status patronage—relative to their own—than lower. This preference for mixing upward rather than downward is present in other works \cite{dong2020segregated, bokanyi2021universal}. There is also a growing body of evidence that urban structure, from barriers to amenities, can distort mobility, reduce intergroup interaction, and limit social ties \cite{abbiasov202415, pinter2025quantifying, toth2021inequality, aiello2025urban}. Yet most of this literature examines discrete urban features—blocks, parcels, activity spaces, or neighbourhood reputations—rather than the planned city as a coherent design project. It shows that people sort in space, but tells us less about how planned units configured into an urban system make that sorting easier, harder, or simply different.

Brasília is a laboratory which we can use to study how different kinds of planned communities, and the system that links them, are implicated in mobility because of its unique history. Conceived as a collection of government buildings—the fuselage—between two wings of housing for government employees, the planned city became part of a larger project, as planned developments and unplanned settlements were forming before construction finished on the Plano Piloto, each with distinct character \cite{costa2019spatial, derntl2019brasilia}. The urban system that emerged gives a unique lens through which to view the interaction of urban environment and human movement. Here we leverage Brasília’s variegated urban structure to understand how the streets, buildings and amenities in a city influence the mobility of its residents. Because its districts were often conceived as replicable planning ``units'' with explicit assumptions \cite{perry1929neighborhood, costa1957planopiloto} about proximity, encounter, and separation, Brasília lets us observe how planning doctrine translates into realized activity spaces—and how those activity spaces do (or do not) connect social groups. 

\begin{figure*}[bt!]
\centering
\includegraphics[width=1\textwidth]{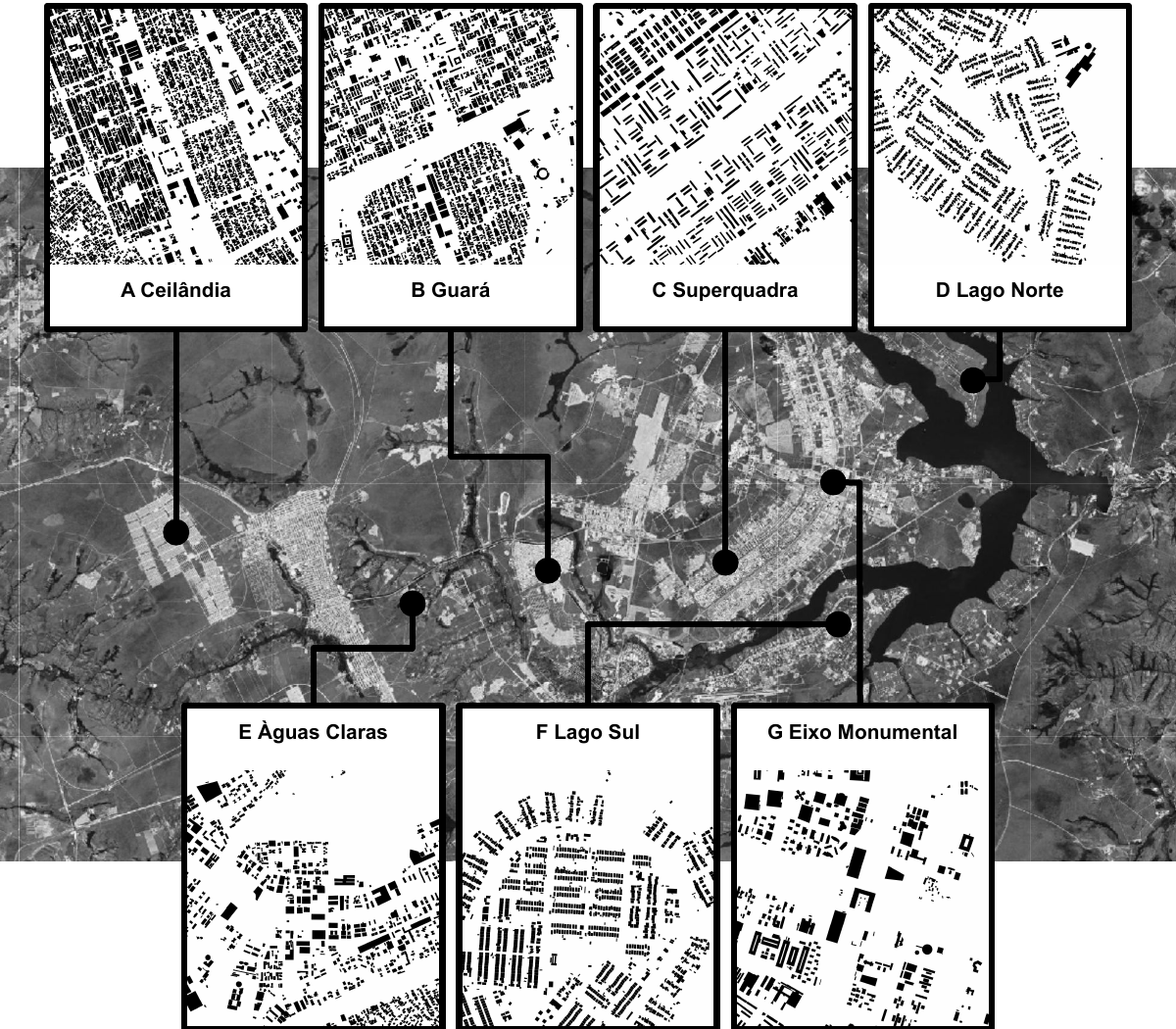}
\caption{\textbf{The structure, form and history of Brasília.} Images taken as part of the Corona program in 1972 showing Brasília 17 years after it broke ground: even before the Plano Piloto was finished, construction had already begun on the Guará suburb and the Ceilândia community for those relocated from informal settlements nearer to the centre. Águas Claras, which was not developed until 1992, is vacant in the image. We also show figure-ground images of different zones as they exist today, with a variety of forms spanning towers-in-the-park superquadras, the dense grids of Guará and Ceilândia and the sparse suburbs of Lago Sul and Lago Norte.}
\label{fig1}
\end{figure*}

\subsection*{Brasília as a metropolis of planned “units”}
Brasília is unusually legible as a planning artifact: the core of the capital was conceived and largely executed within a single presidential term, and the city was then forced to metabolise rapid growth through a succession of later planned settlements \cite{brasil1956novacap, costa1957planopiloto, unesco2026brasilia}. (Note that to add necessary context to our analysis we put short histories of each satellite city in Brasília in Supplementary Section ~\ref{histories}.) Because of this compressed history, the city reads less as one continuous fabric than as an accumulation of explicit planning ideas, preserved clearly enough that their seams remain visible \cite{iphan2016patrimonio}. The core ``Plano Piloto'' emerged from the 1957 competition won by Lúcio Costa, with the emblematic civic architecture entrusted to Oscar Niemeyer and the work of urbanisation coordinated by NOVACAP; in the official language of the period, the capital was imagined not simply as a container for federal government but as an instrument of national modernisation \cite{brasil1956novacap, costa1957planopiloto, unesco2026brasilia}. Brasília as it was built was therefore always both a diagram and a settlement, both a plan and a process. 

Within the Plano Piloto, Costa’s logic is modular and scalar. The 1957 report lays out the crossed axes, the concentration of civic and administrative functions along the monumental axis, and the long residential wings arranged along the curved road axis \cite{costa1957planopiloto}. In Brasília Revisitada, Costa later names four ``urban scales'', or \emph{escalas}: monumental, residencial, gregária, and bucólica \cite{costa1987revisitada}. In his account, escala gregária is the denser central layer of commerce, services, leisure, and exchange, meant to give thickness to everyday urban life in counterpoint to the more open residential wings \cite{costa1987revisitada}. In these wings, the superquadra is the key residential unit: a repeated sequence of apartment slabs along the residential axis, shown in Fig. ~\ref{fig1}\textbf{C}, with entrequadras holding schools, churches, local commerce, and shared facilities \cite{costa1957planopiloto, iphan2025superquadra}. Costa also describes an intended public sphere along the ground plane—residential blocks on pilotis, open green space, and continuous pedestrian routes below—while making clear how much the circulation system was designed around the automobile: Brasília was planned to facilitate vehicular flow through a hierarchical road network while concentrating pedestrian life within the superquadras \cite{costa1957planopiloto}. The plan also assumed a strong municipal bus service using these circulation corridors for collective transport, a dimension Costa later noted had not been satisfactorily realized \cite{costa1987revisitada}. 

Around Lake Paranoá, low density residential subdivisions took shape in Lago Sul and Lago Norte; their sparse curving layouts, evident in Figs. ~\ref{fig1}\textbf{D} and \textbf{F}, around the water differ from both the linear slabs of the Plano Piloto and the denser grids to the west \cite{costa1987revisitada, manicoba2019regioes}. Even here, however, Costa insisted that the orla, or shore, should remain public and accessible rather than become a private residential amenity \cite{costa1987revisitada}.

But the city did not remain confined to the diagrammatic ideal for long. As the 1972 Corona image in Fig. ~\ref{fig1} already suggests, metropolitan Brasília was taking shape beyond the Plano Piloto even before the planned centre had fully matured. Early work on Brasília showed that the planned city and the city that was actually built diverged almost immediately: the Plano Piloto did not provide adequate housing for many of the workers and migrants drawn in by construction, so labour camps, informal settlements, and satellite communities grew alongside the official capital from the outset \cite{epstein1973plan}. Because its formal housing supply could not absorb the scale and speed of population growth around the new capital, early policy turned to the removal of occupations and the transfer of residents to satellite settlements, producing from the outset a multinucleated urban region \cite{ipea2015governanca, costa2016taguatinga}. Taguatinga, founded in 1958, was the first official satellite city and set an early template for peripheral growth management through planned relocation \cite{costa2016taguatinga}.

Several of these later settlements were themselves planned as distinct nuclei, with their own sectors, residential units, and service areas \cite{codeplan2018guara, codeplan2018ceilandia}. In particular, two large planned suburbs, Guará and Ceilândia, emerged with distinct origins and planning rationales \cite{manicoba2019regioes}. With construction beginning in 1967, Guará I was conceived as housing for workers, especially NOVACAP employees, and Guará II, built in 1972, extended that role as housing for public employees \cite{manicoba2019regioes, codeplan2014guara}. Ceilândia, by contrast, was founded in 1971 under the Campanha de Erradicação de Invasões, CEI, as a planned resettlement city north of Taguatinga \cite{codeplan2018ceilandia, manicoba2019regioes}. Here, invasões is the state’s term for informal occupations; the name ``Ceilândia'' itself, CEI + lândia, preserves the administrative logic of that early resettlement process \cite{manicoba2019regioes}. The initial plan established an urban area of 20 km$^2$ with roughly 17,000 lots for the transfer of residents removed from informal settlements near the core \cite{codeplan2018ceilandia}. Placed side by side in Fig. ~\ref{fig1}\textbf{A} and \textbf{B}, Guará and Ceilândia differ from the superquadras morphologically, yet they are no less planned. They translate the same ethos of ordered settlement into denser grids and smaller parcels \cite{codeplan2018guara, codeplan2018ceilandia}.  

Urban life in these relocated communities diverged from the modernist communities of the planned city. Within the Plano Piloto, residents describe the modernist centre as green and pleasant while lacking the street life and spontaneous sociability associated with older cities \cite{holston1989modernist}. Interviews with residents from its satellites focus on poor infrastructure and services, but also, and as a consequence, networks of mutual aid and support that strengthened social life in these new towns \cite{derntl2024capitality}. 

Situated between the Guarás and Ceilândia is the newer Águas Claras, developed by Paulo de Melo Zimbres, who presented the design as an attempt to recover something closer to the traditional street \cite{pereira2021quarteirao,williams2007brasilia}. A 1992 law authorised the district’s construction, with early plans anticipating buildings of up to 12 stories, but implementation produced a much more intense verticalisation than the original conception suggested \cite{distritofederal1992aguasclaras, pereira2021quarteirao, ipedf2025verticalizacao}. The resulting fabric is a high-rise, mixed-use district of condominium blocks, commercial strips, and selective active frontages \cite{codeplan2019aguasclaras, pereira2021quarteirao}. 

These communities have scaffolded the evolution of the broader city: residence has spread across multiple planned nuclei while employment remains strongly concentrated in the Plano Piloto \cite{codeplan2020trabalho}. The result is not simply metropolitan growth but metropolitan asymmetry: the capital grows outward even as daily work still bends inward \cite{codeplan2020trabalho, goncalves2024deslocamento}. That asymmetry lengthens commutes and intensifies pressure on the transport system \cite{goncalves2024deslocamento}. A survey of residents found that of roughly 1.2 million employed commuters, the Plano Piloto was the primary destination, drawing about 41.6\% of commuters and generating more than 514 thousand daily work trips \cite{goncalves2024deslocamento}. The same study notes that the Plano Piloto draws disproportionately higher income and more educated workers, and proportionally fewer Black workers \cite{goncalves2024deslocamento}. By the present day, then, the key contrast is no longer only between plan and deviation, but among several planning logics now interlocked through one metropolitan labour market.

\noindent Rather than treating Brasília only as an iconic modernist centre, we treat it as a metropolitan system of planned units, unequal centralities, and hard discontinuities. This lets us ask three linked questions with direct planning relevance: where do different income groups actually mix in everyday mobility, which groups bear the travel burden when such mixing occurs, and which spatial conditions---especially amenities, barriers, and neighbourhood form---make cross-class encounter more or less likely. Using anonymised GPS traces, a street-bounded tessellation, and a classification of urban form, we show that mixing is concentrated in a small set of amenity-rich centralities, that poorer residents travel farther to produce it, and that barriers and enclosure structure help turn short metric distances into durable social separation. The contribution is therefore not only to the literature on experienced segregation, but to a planning question at the heart of Brasília and other planned capitals: how urban structure distributes access to shared urban life.

Because of this, Brasília offers a diverse menu of urban characters within a single metropolitan field: high modernism in the Plano Piloto, planned satellites in Guará I and II, resettlement urbanism in Ceilândia, low density residential sectors around the lake, and a corridor of towers in Águas Claras \cite{costa1957planopiloto, codeplan2018guara, codeplan2018ceilandia, pereira2021quarteirao}. These communities do not just represent aesthetic differences. They encode distinct assumptions about residence, circulation, access, and the organisation of daily life \cite{costa1957planopiloto, costa1987revisitada}. Seen this way, Brasília is not a single ``case'' but a bundle of comparisons: one metropolitan area, several urban forms, several access regimes \cite{ipea2015governanca, codeplan2018amb}. This creates an unusually clean setting to ask whether urban form and function shape who meets whom, or simply rearrange where segregation is experienced \cite{browning2022geographic}. The following paper investigates day-to-day human mobility across the constellation of cities that constitute Brasília through the lens of urban structure—the buildings, roads and amenities that shape urban life. We first define a morphometric classification system for the city’s residential neighbourhoods, commercial corridors and open spaces. We then use GPS mobility data to quantify differences in interaction patterns across those urban forms and functions, finding clear patterns of intergroup interaction and ``compelled mobility'' \cite{browning2022geographic}, or who moves to whom when social mixing happens.

\section*{Methods}
This study combines three datasets covering the Brasília metropolitan area: GPS-based mobility traces, data on urban form and function, and census income records.

\paragraph{Mobility data.} Anonymised GPS traces were obtained from Locomizer for a one-month observation period, comprising 27.4 million stays by 331,934 unique users. Stays were detected using a tessellation, which we will describe later, with 30-minute time slots, using an established algorithm \cite{aslak2020infostop}. Each user's income was estimated from their residential location and associated metadata. Users were assigned to one of five quantiles on the logarithm of income, yielding approximately 20\% of users in each group. Radius of gyration is summarised for users with at least 10 stays. Stratified analyses of individual segregation $S_i$ use a subset of users with at least 10 stays, reducing instability from sparse traces.

Coverage and bias checks for the GPS panel are reported in Supplementary Fig. ~\ref{data_validation}. Users span the full spatial extent, and the remaining socio-demographic biases are modest relative to the segregation gradients analysed below. The slight bias towards wealthier residents cuts \emph{against} our core findings, so our estimates are more likely to be conservative. 

\paragraph{Built and natural environment.} Building footprints (208,758 structures) and road network data were sourced from the Overture Maps Foundation \cite{}. We used these buildings to construct ``enclosed tessellations'' \cite{arribas2022spatial}: this approach computes a Voronoi cell for each building, clipped by the street network such that no cell crosses a road. This produced 218,609 tessellation cells---207,887 associated with buildings and 10,722 gap cells in interstitial spaces---covering 1{,}671\,km\textsuperscript{2}. The approach also naturally produces a hierarchy of enclosures, according to various levels of division---blocks produced by all streets, neighbourhoods produced by secondary roads, and districts produced by primary roads, none of which cross barriers like water areas or train tracks. 

\paragraph{Socio-economic data.} Per capita income data from the 2010 Brazilian Census were spatially interpolated from 4{,}454 census tracts to tessellation cells via centroid-based spatial joins, achieving 98.2\% coverage.

\subsection*{Measuring Segregation}
Borrowing from earlier work \cite{moro2021mobility}, we adopt an approach that quantifies segregation at two complementary scales: the place and the individual. \textbf{Place segregation} ($S_\alpha$) measures how evenly visitors to a given location $\alpha$ are distributed across income groups. For $N = 5$ income quintiles, let $\nu_{q\alpha}$ denote the proportion of unique visitors to cell $\alpha$ belonging to quintile $q$. The place segregation index is specified as

\begin{equation}
S_\alpha = \frac{N}{2(N-1)} \sum_{q=1}^{N} \left| \nu_{q\alpha} - \frac{1}{N} \right|
\label{eq:salpha}
\end{equation}

\noindent where $S_\alpha$ ranges from 0 (perfect mixing: all quintiles equally represented) to 1 (complete segregation: visitors drawn from a single quintile). To ensure robust estimation, $S_\alpha$ is computed only for cells receiving at least 10 unique visitors, yielding 26{,}453 cells (12.1\% of the tessellation) with a mean of 0.498 (s.d.\ = 0.200).

\textbf{Individual segregation} ($S_i$) captures the segregation experienced by a specific person $i$ across all the places they visit. Let $w_{i\alpha}$ denote the share of person $i$'s visits that occur in cell $\alpha$, and $\nu_{q\alpha}$ the income composition of that cell as above. The exposure of individual $i$ to quintile $q$ is

\begin{equation}
E_{qi} = \sum_{\alpha} w_{i\alpha} \, \nu_{q\alpha}
\label{eq:exposure}
\end{equation}

\noindent and the individual segregation index is

\begin{equation}
S_i = \frac{N}{2(N-1)} \sum_{q=1}^{N} \left| E_{qi} - \frac{1}{N} \right|
\label{eq:si}
\end{equation}

\noindent with $S_i$ computed for individuals with more than 10 stays, then aggregated to the cell level by averaging over residents in each home cell, producing mean $S_i$ values for 5{,}307 cells.

The two indices are complementary: $S_\alpha$ characterises a \emph{place} irrespective of who visits it, while $S_i$ characterises a \emph{person's} experienced environment across all their daily activities. Their difference, $\Delta_i = S_i - S_{\alpha,\text{home}}$, captures the extent to which individual mobility mitigates or reinforces the segregation of one's home neighbourhood.

Sensitivity to the minimum-visitor and minimum-observation thresholds used for $S_\alpha$ and $S_i$ is reported in Supplementary Figs. ~\ref{threshold_sensitivity} and ~\ref{si_threshold_sensitivity}. These filters shift reported levels somewhat, often for mechanical reasons because we are selecting on an individual characteristic that may mechanically affect mixing—namely, how much people move—but they leave the broad geography and ordering of segregation intact.

\subsection*{Measuring urban form and function}
For each tessellation cell, we compute a comprehensive suite of morphometric conditions \cite{fleischmann2019momepy}. These quantify the geometric and relational properties of buildings (area, elongation, compactness, convexity, orientation), cells (area, perimeter, compactness), building--cell relationships (coverage ratio, open space ratio), the street network (density, connectivity, centrality at local, district, and city scales), and the enclosure hierarchy (area, cell count). Raster-derived measures contribute building height, volume, surface area, population density, and NDVI \cite{}. Amenity diversity is captured through richness, evenness, and categorical counts of points of interest.

Each primary character is supplemented by contextual measures---the interquartile mean (IQM) and interquartile range (IQR)---computed over a spatial lag of 10 topological steps using Queen contiguity weights with inverse-distance weighting. This captures the morphological context surrounding each cell: the IQM represents the typical value in the neighbourhood while the IQR captures local heterogeneity. After removing zero-variance and redundant columns, the final feature set comprises 121 standardised variables.

\paragraph{Spatial signatures.} To classify the urban fabric into distinct morphotypes, we apply $k$-means clustering ($k = 12$, $n_\text{init} = 500$) to the standardised feature matrix, following the spatial signatures framework of \cite{arribas2022spatial}. The optimal $k$ was selected via diagnostic clustergram analysis across $k = 2$--$25$, evaluating silhouette scores, Calinski--Harabasz and Davies--Bouldin indices, and inertia. The resulting 12 types range from \emph{civic core} and \emph{superquadra} (Brasília's iconic modernist residential blocks) to \emph{dense, poor, irregular} and \emph{dense, poor, regular}, distinguished by street and building alignments, \emph{emergent suburb}, and \emph{fringe suburb}, each characterised by a distinctive morphological profile. The classification is derived from urban form, street topology, amenities, and parcel qualities like size, coverage and vegetation.

Distributional summaries of the variables that define these morphotypes are reported in Supplementary Figs. ~\ref{morphotype_violins_1} and ~\ref{morphotype_violins_2}. These supplementary plots show that the superquadra is not simply a wealthier residential type, but a distinctive combination of elongated slabs, low coverage, greenness, and high income. They also demonstrate the difference between irregular and regular settlements---which show distinct street alignments---and show that very few morphometric classes have amenities, which is relevant to our study of mixing because it limits the spaces where interaction might be facilitated. 

\subsection*{Modelling Framework}
The morphometrics give us 122 features in total, which present dimensionality and collinearity problems. To identify which morphological features predict place segregation and visitor volume, we employ Elastic Net regression---a regularised linear model that combines the sparsity-inducing properties of the LASSO (L1 penalty) with the collinearity-handling capacity of Ridge regression (L2 penalty). Given the high dimensionality (121 features) and substantial multicollinearity among morphometric variables (e.g., building area, surface, and volume), standard OLS would produce unstable coefficient estimates; Elastic Net addresses both problems simultaneously. Given a response vector $\mathbf{y}$ and a standardised feature matrix $\mathbf{X}$ of $n$ observations and $p$ features, the Elastic Net minimises:

\begin{equation}
\hat{\boldsymbol{\beta}} = \underset{\boldsymbol{\beta}}{\arg\min} \left\{ \frac{1}{2n} \| \mathbf{y} - \mathbf{X}\boldsymbol{\beta} \|_2^2 \;+\; \alpha \left[ \rho \| \boldsymbol{\beta} \|_1 \;+\; \frac{1-\rho}{2} \| \boldsymbol{\beta} \|_2^2 \right] \right\}
\label{eq:elasticnet}
\end{equation}

\noindent where $\alpha > 0$ controls the overall strength of regularisation and $\rho \in [0, 1]$ (the L1 ratio) governs the balance between the L1 and L2 penalties. When $\rho = 1$ the model reduces to the LASSO, which drives weak coefficients to exactly zero, performing automatic feature selection. When $\rho = 0$ it reduces to Ridge regression, which shrinks correlated features toward each other rather than selecting one arbitrarily. Intermediate values of $\rho$ yield a compromise: groups of correlated features are retained but with reduced coefficients, while uninformative features are eliminated.

Both $\alpha$ and $\rho$ are selected via five-fold cross-validation, searching over $\rho \in \{0.1, 0.3, 0.5, 0.7, 0.9, 0.95, 0.99\}$ and 100 values of $\alpha$ on a logarithmic grid. All 121 features are standardised to zero mean and unit variance prior to fitting, so that the resulting coefficients $\hat{\beta}_j$ are directly comparable as standardised effect sizes: a one standard deviation increase in feature $j$ is associated with a $\hat{\beta}_j$ change in the dependent variable. We estimate three specifications:

\begin{enumerate}
    \item \textbf{DV1}: $\log(\text{visitors}_\alpha) \sim \mathbf{X}$, identifying which urban features attract higher visitor volumes;
    \item \textbf{DV2}: $S_\alpha \sim \mathbf{X}$, identifying which features predict place segregation;
    \item \textbf{DV3}: $S_\alpha \sim \mathbf{X} + \log(\text{visitors}_\alpha)$, a mediation model testing whether the effect of morphology on segregation operates partly through visitor volume.
\end{enumerate}

\noindent Comparing DV2 and DV3 reveals the extent to which morphological predictors of segregation act directly---through the built form's influence on who encounters whom---versus indirectly, by shaping how many people visit a location.

\section*{Results}
One metropolitan asymmetry should be kept in view from the outset. Amenities and visits cluster around the Plano Piloto and adjacent centralities, whereas population is distributed much more broadly across satellite cities and peripheral settlements (see Supplementary Fig. ~\ref{spatial_context}). Brasília therefore combines decentralised residence with centralised opportunity. That mismatch is the structural backdrop of the results below: mobility can widen exposure, but only through a metropolitan landscape in which the most socially integrative destinations are spatially concentrated.

\begin{figure*}[bt!]
\centering
\includegraphics[width=1\textwidth]{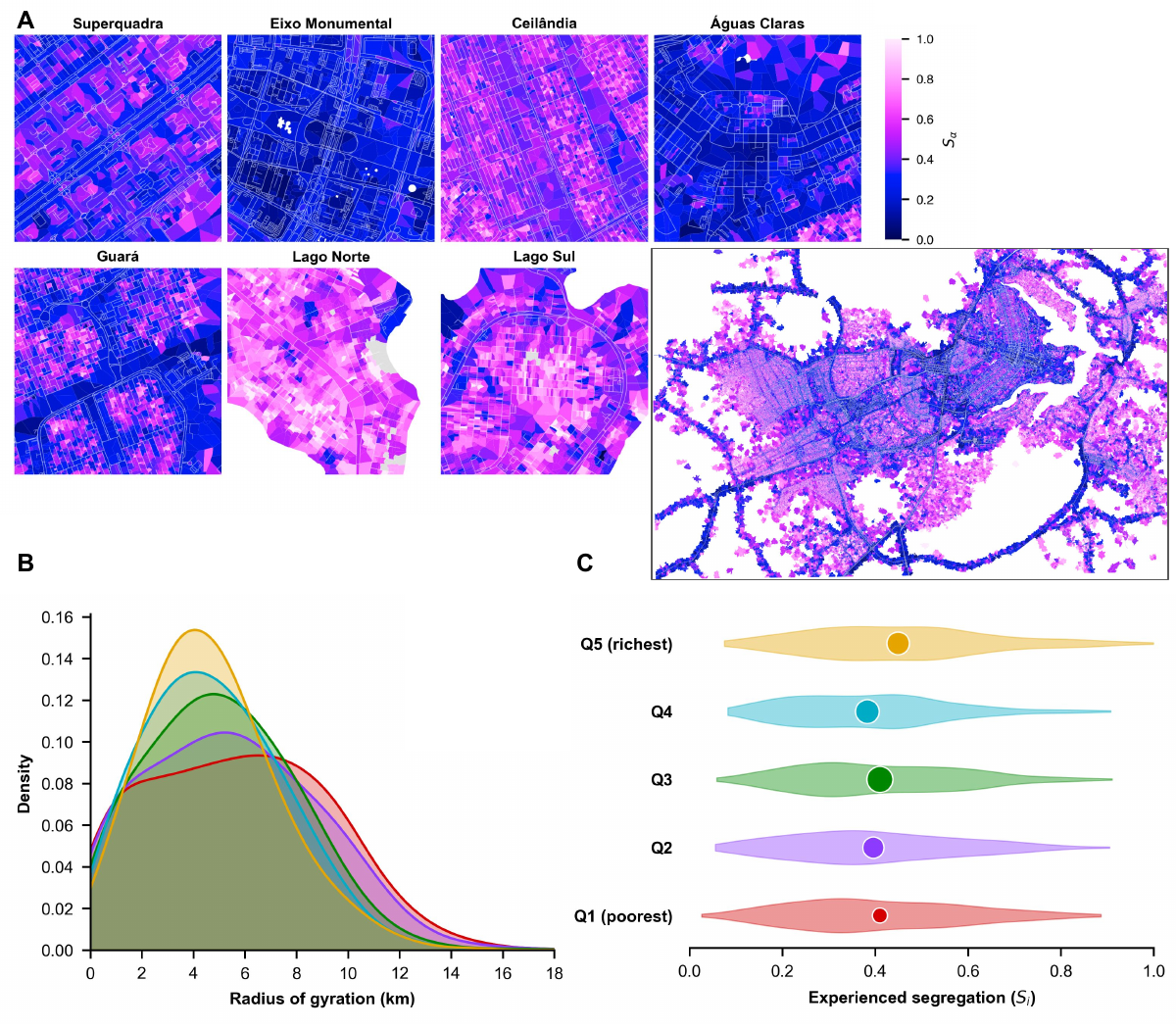}
\caption{\textbf{Experienced segregation in Brasília.} \textbf{A} Place segregation by tessellation in selected areas. Lago Sul and Lago Norte are segregated almost throughout, while the superquadras and Guará often combine segregated residential cells with more mixed commercial corridors. \textbf{B} Radius of gyration by income quintile. Poorer groups travel farther each day than richer ones, consistent with compelled mobility. \textbf{C} Individual experienced segregation $S_i$ by income quintile. Despite travelling less, the richest quintile remains the most segregated in experience.}
\label{fig2}
\end{figure*}

We first compute place segregation $S_\alpha$ on a tessellation bounded by streets and other hard barriers. At the street-and-building scale, active places span almost the full range of the index, from $0.028$ to $1.000$ (median $=0.487$). Fig. ~\ref{fig2}\textbf{A} shows that this variation is intensely local: mixed commercial corridors sit beside segregated residential tissue, especially in the superquadras and in Guará. The same pattern strengthens under aggregation. Weighted mean $S_\alpha$ rises from $0.282$ at the district scale to $0.357$ at the neighbourhood scale and $0.545$ at the block scale, indicating that Brasília looks most integrated when viewed coarsely and most segregated at the scales closest to everyday access and encounter. The lowest values appear in the major commercial concentrations around Águas Claras and along the Monumental Axis; the highest occur in the lakefront suburbs, whose network isolation and water boundaries limit broader encounter. 

These differences are stable across reasonable threshold choices and are not an artefact of sparse sampling, as our sensitivity analysis demonstrates in Supplementary Fig. ~\ref{threshold_sensitivity}. Nor are they a trivial byproduct of activity volumes alone: a null model that maintains individual activity levels and cell popularity, while preserving the distance distribution, still produces a much flatter segregation landscape than the observed one. (See Supplementary Fig. ~\ref{null_model_segregation} for more detail.)

Mobility softens these residential divides, but it does not erase them. Across our sample, 77.2\% have $S_i < S_{\alpha,\mathrm{home}}$, with a median reduction of 0.119 and a Pearson correlation of 0.473 between the two measures. Fig. \ref{fig2}\textbf{B} shows the variation in radius of gyration---a measure of how far people move each day---per income quintile, and documents an important difference between rich and poor: the poorest groups move farther each day than the wealthiest groups. Among users with at least 10 stays, the poorest quintile has the largest median radius of gyration ($R_g=4.18$~km), compared with $3.98$~km for the richest quintile. Yet the richest quintile remains the most segregated in experience (median $S_i=0.431$, versus $0.373$ for the poorest quintile), shown in Fig. ~\ref{fig2}\textbf{C}. Alongside this stylised fact, we see in Fig. \ref{fig2}\textbf{C} that this difference in aggregate mobility coincides with greater individual segregation $S_i$ among the richest quintile than the other quintiles, which are generally similar. 

The point is clearest in the exposure vectors themselves. The richest quintile's average exposure to other rich users is 0.521, while its exposure to the poorest quintile is only 0.091. For the poorest quintile, within-class exposure is lower, at 0.440, and between-class exposure is broader. What distinguishes affluent mobility in Brasília is therefore not just how far people move, but the social composition of the places they reach. Mobility for poorer groups is more expansive and more heterogeneous; mobility for richer groups is more selective. The city thus widens activity spaces without equalising them: poorer residents travel farther, while affluent residents retain more socially selective exposure. This pattern is consistent with ``compelled mobility'' \cite{browning2022geographic}, wherein the residents may be forced to move more or farther because key resources are missing locally.  

This widening of activity space is also unequal in its costs. When cross-class co-presence occurs, poorer residents are more likely to do the travelling required to produce it. The poorest quintile travels 2.87 times farther than the richest to reach mixed-class encounters, and the asymmetry is especially stark in Brasília's planned central morphotypes, where poor-to-rich mixing requires substantially more movement than the reverse (see Supplementary Fig. ~\ref{burden_of_mixing}). Brasília therefore does not mix symmetrically: mobility softens segregation for many users, but the burden of that softening falls disproportionately on poorer groups.

\subsection*{Experienced segregation varies systematically across urban forms}
\begin{figure*}[bt!]
\centering
\includegraphics[width=1\textwidth]{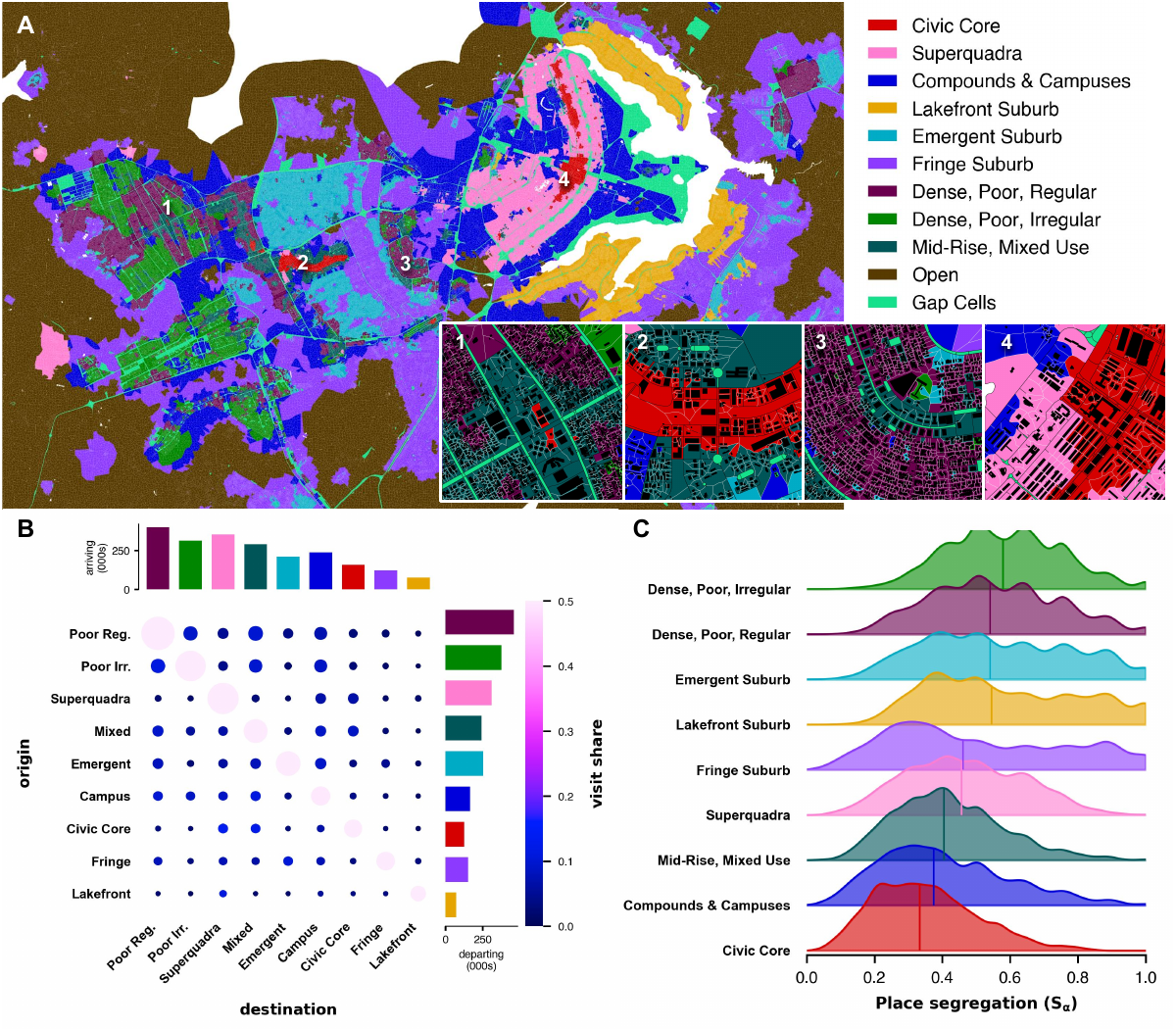}
\caption{\textbf{Morphotypes and experienced segregation.} \textbf{A} The results of our morphometric classification, which distinguishes between commercial zones in red and various residential communities, including the superquadras within the Plano Piloto in pink, Lago Sul and Lago Norte in yellow, Ceilândia in purple and green. \textbf{B} The matrix of interactions between different morphotypes, showing that mobility concentrates within morphotypes---when trips cross morphotypes, residential areas feature as sources and commercial corridors as sinks. \textbf{C} Place segregation $S_\alpha$ across morphotypes, showing that civic cores---which are present in both the Plano Piloto and Águas Claras---have the lowest place segregation; suburbs typically have higher place segregation.}
\label{fig3}
\end{figure*}

We then classify each cell according to its morphological properties, its topological location in the city, and other contextual features. The results of the clustering are mapped in Fig. ~\ref{fig3}\textbf{A}. The city shows two primary central places, ``civic cores'' that are rich in amenities; the morphometric classification also identifies the modernist superquadras of the Plano Piloto, as well as the lakefront suburbs of Lago Sul and Lago Norte. In addition to distinguishing morphotypes of planned satellite suburbs such as Guará and Ceilândia, the clustering reveals another class that we term emergent suburbs, which typically correspond to areas shaped by \emph{parcelamentos irregulares}---land subdivided and settled incrementally before consolidating into coherent suburban districts. We also find mid-rise, mixed-use areas in Guará, Ceilândia and Águas Claras: areas with a combination of residential and commercial uses; these areas are often nearby civic cores or major roads. Another class demarcates open areas with large buildings and commercial or institutional functions, including hospitals, embassies, clubs, military campuses, as well as large stores, restaurants and hotels along major roads; we call this class ``Compounds \& Campuses''. Fig. ~\ref{fig3}\textbf{B} shows the spatial interaction patterns between these morphotypes, with the strong diagonal demonstrating that most trips occur within a given morphotype and that most activity occurs in the residential communities, not the commercial corridors. 

We begin with a simple question: does experienced segregation vary systematically with urban form? Morphotypic comparison makes clear that Brasília does not exhibit a single exposure effect. We show the distributions of place segregation by morphotype in Fig. ~\ref{fig3}\textbf{C}. Although they account for comparably fewer visits than other areas, civic cores are the least segregated environments (median $S_{\alpha}=0.333$), followed by mid-rise, mixed-use tissues (0.405) and superquadras (0.457). At the segregated end sit dense, poor, regular areas (0.542), Emergent Suburbs (0.542), Lakefront Suburbs (0.545), and Dense Poor Irregular areas (0.580). One result is especially revealing: Dense, poor areas and the rich suburbs along the lake do not exhibit statistically distinguishable levels of place segregation. The city's poorest irregular fabrics and its wealthiest enclaves therefore arrive at similar levels of place segregation, albeit through different spatial logics: Lago Sul and Lago Norte are close to the city centre as the crow flies, but removed from it in the street topology; many of the denser, poorer parts of the city are at the fringe both topologically and geographically.

A grouped comparison sharpens the ranking. Mixed fabrics like the civic core and mid-rise, mixed-use typologies have a median $S_{\alpha}$ of 0.390, compared with 0.477 for all suburbs and 0.558 for both dense, poor tissues. Yet these mixed areas are not well distributed across the metropolis: citywide, civic core cells make up just 0.7\% of the classification and mid-rise, mixed-use cells make up 3.6\%. Brasília therefore does not lack integrative space in principle; it lacks enough of it to counterbalance the much larger field of socially narrower residential environments.

A temporal split clarifies what kind of mixing these centralities sustain; we compute separate indices for weekdays and weekends in Supplementary Fig. ~\ref{weekday_weekend_segregation}. On weekdays, they draw broader publics through work, services, and circulation; on weekends, they become more segregated, with the largest shifts in civic core and superquadra. Some of Brasília's most integrated spaces become less mixed settings without commutes driving mobility to these areas.

\begin{figure*}[bt!]
\centering
\includegraphics[width=1\textwidth]{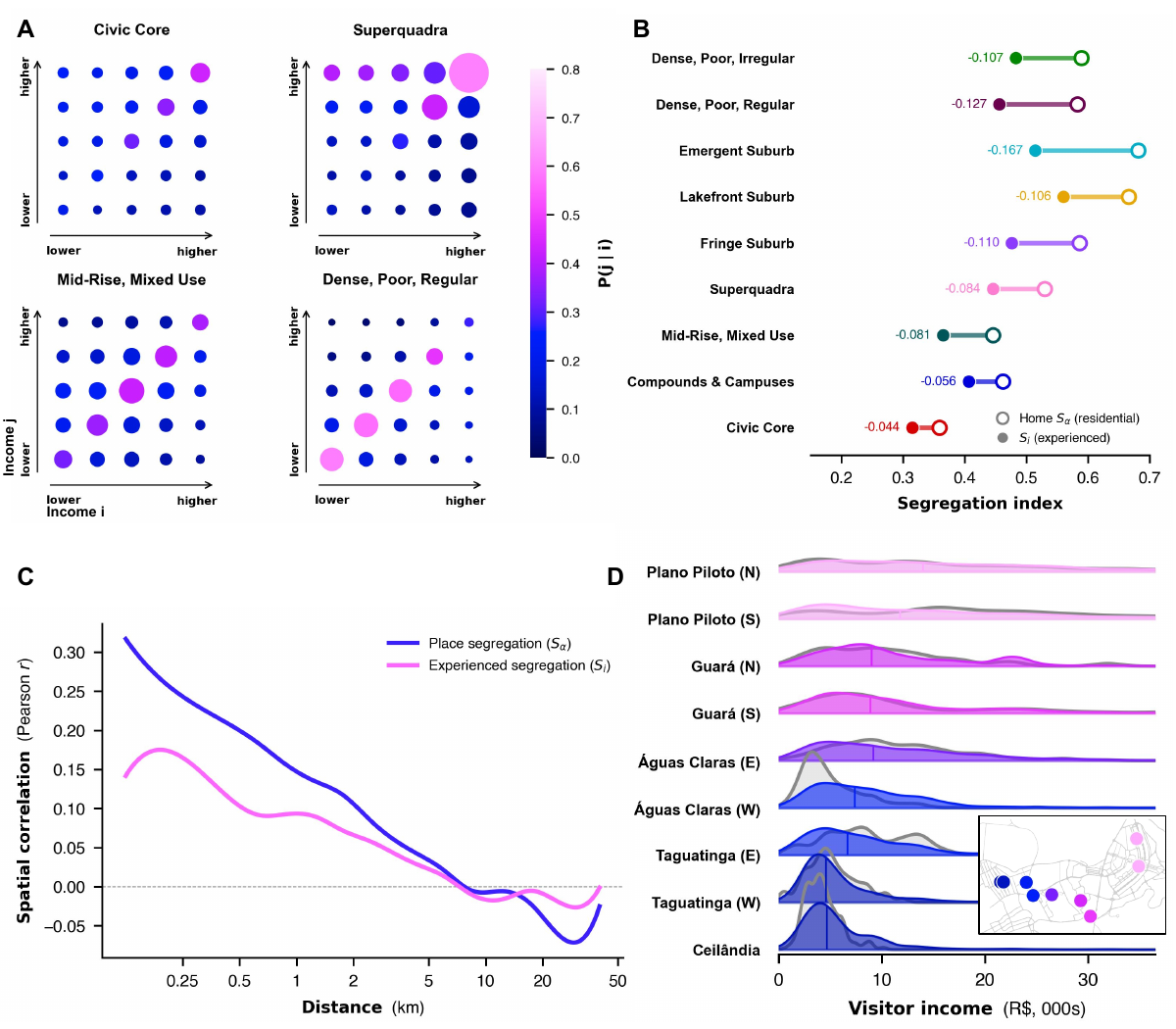}
\caption{\textbf{Who interacts with whom and where.} \textbf{A} Mixing matrices for select morphotypes show the probability that a user from income group $i$ shares a cell and time window with a user from group $j$. The diagonals mark assortativity: across all morphotypes, within-class encounter is more likely than between-class encounter, but the composition of that assortativity varies sharply across urban forms. \textbf{B} Individual experienced segregation $S_i$, grouped by home morphotype. Mobility usually reduces segregation relative to the home cell, but the extent of that reduction varies by residential setting. \textbf{C} Spatial correlograms for place segregation $S_\alpha$ and resident experienced segregation $S_i$. Nearby places resemble one another more strongly than nearby residents do. \textbf{D} Visitor income distributions for contiguous civic-core clusters. Similar centralities attract different publics depending on their position in the metropolitan system.}
\label{fig4}
\end{figure*}

As we saw above, mobility between morphotypes is not random: visits are concentrated within morphotypes, but when trips do cross morphotype boundaries, residential areas act primarily as sources and the commercial ones as sinks. The same asymmetry appears in the mixing matrices of Fig. ~\ref{fig4}\textbf{A}, which shows the probability that different groups interact with each other \emph{conditioning on morphotype}: within morphotypes, contact is assortative, but the composition of that contact shifts systematically across the city. In superquadras, the modal interactions occur among the upper quintiles, whereas in dense poor areas the most common encounters are among poorer groups. Mixed centralities sit between these poles, hosting broader exposure profiles rather than eliminating class structure altogether. Yet although many of these morphotypes show a strong diagonal, indicating homophily even when different groups visit the same morphotype, in superquadras, many lower-class and upper-class visitors still interact.   

While $77.2\%$ of users have $S_i < S_{\alpha}^{\mathrm{home}}$, with a median difference of $-0.119$, we document variation across morphotypes. This softening is strongest for some suburban morphotypes and weakest for already insulated residential types. In the full sample by home morphotype, residents in civic cores have the lowest median experienced segregation ($S_i=0.299$), while residents of lakefront suburbs remain the most isolated (0.500). Residents of superquadra present another important case: despite living in a morphotype that is less segregated than the dense, poor tissues by $S_\alpha$, their average exposure remains tilted toward the affluent (mean exposure to $Q5 = 0.439$ versus $0.096$ to $Q1$).

This sorting is structured not only by distance, but by discontinuity. Place segregation is more spatially autocorrelated than experienced segregation at short distances ($r=0.317$ against $0.142$ within $0$--$0.2$~km; Fig. ~\ref{fig4}\textbf{C}). Yet spatial proximity systematically misrepresents social relationships: when distance is measured along the road network rather than with Euclidean distance, nearby cells appear more socially similar at every distance band, indicating that apparently close places can remain functionally far apart. Matched cell pairs separated by water, open land, or other hard breaks are 20.5\% more dissimilar in income composition than comparable pairs at the same Euclidean distance , and 57.2\% of the variance in place segregation sits at or above the block level of the enclosure hierarchy. The same pattern appears one scale down inside neighbourhoods: cells fronting roads are relatively mixed, but segregation rises sharply just one cell inward and then plateaus. This has strong implications for planning: Brasília's corridors behave as mixing edges, while the interiors they enclose behave as socially narrower cores. Brasília's barriers do not merely slow movement; they partition fields of encounter. (See Supplementary Fig. ~\ref{barriers_matter} for more detail.)

The morphotypes are not only formal descriptors. They are socially differentiated residential worlds. Home morphotype and income quintile are strongly associated ($\chi^2 = 5240.2$, $p < 0.001$, Cramer's $V = 0.384$). Among the richest quintile, $50.6\%$ live in superquadras and another $12.8\%$ in lakefront suburbs. Among the poorest quintile, $64.1\%$ live in dense, poor, regular or irregular tissues. Read the other way, $94.8\%$ of lakefront-suburb residents belong to the top quintile, while just $0.3\%$ of dense poor informal residents do. Before anyone in Brasília leaves home in the morning, the city has already sorted classes into different urban settings. To show that mixing is relational rather than absolute, we disaggregate the city's civic cores into contiguous clusters in Fig. ~\ref{fig4}\textbf{D}. Similar centralities attract different publics depending on metropolitan position: the northern side of the Plano Piloto receives a wealthier visitor profile than the southern side, and eastern Águas Claras a wealthier one than its western half. Exposure begets mixing, but the social content of that exposure depends on which basin a node sits in.

\subsection*{Form and function explain where mixing does and does not emerge}
To isolate which features matter most, we estimate elastic net models using 121 standardised urban, demographic, and contextual predictors. We fit three specifications: one for visitor volume, one for place segregation $S_\alpha$, and a third that adds $\log(\mathrm{visitors})$ to the segregation model as a mediation test. The structure of the models is informative in its own right. Full model summaries and supplementary diagnostics are reported in Supplementary Section ~\ref{models}. They reinforce the same qualitative point: morphology explains far more of segregation than visit volume alone. The visitation model behaves like a Ridge regression ($L_1$ ratio $= 0.10$) and retains 82 predictors, implying that footfall is assembled from many correlated ingredients. The segregation models are far sharper: both are almost pure LASSO fits ($L_1$ ratio $= 0.99$), retaining 61 and 66 terms. Making a place busy therefore looks additive. Making it socially mixed is more selective.

Model comparison sharpens the planning interpretation. Urban form and function explain visitation better than segregation: the model for $\log(\mathrm{visits})$ reaches $R^2=0.436$, whereas the model for place segregation reaches $R^2=0.363$ using urban features alone and $R^2=0.378$ after adding visitor volume. The modest improvement from adding $\log(\mathrm{visits})$ ($\Delta R^2 = 0.015$) indicates that busy places tend to be less segregated, but crowding alone does not explain why some places mix and others do not. The contrast is even clearer when visitor volume is used on its own: $\log(\mathrm{visitors})$ explains only 14.9\% of the variance in $S_{\alpha}$, far below the 37.8\% explained once urban features are reintroduced. Crowds matter, but they do not settle the question. 

\begin{figure*}[bt!]
\centering
\includegraphics[width=1\textwidth]{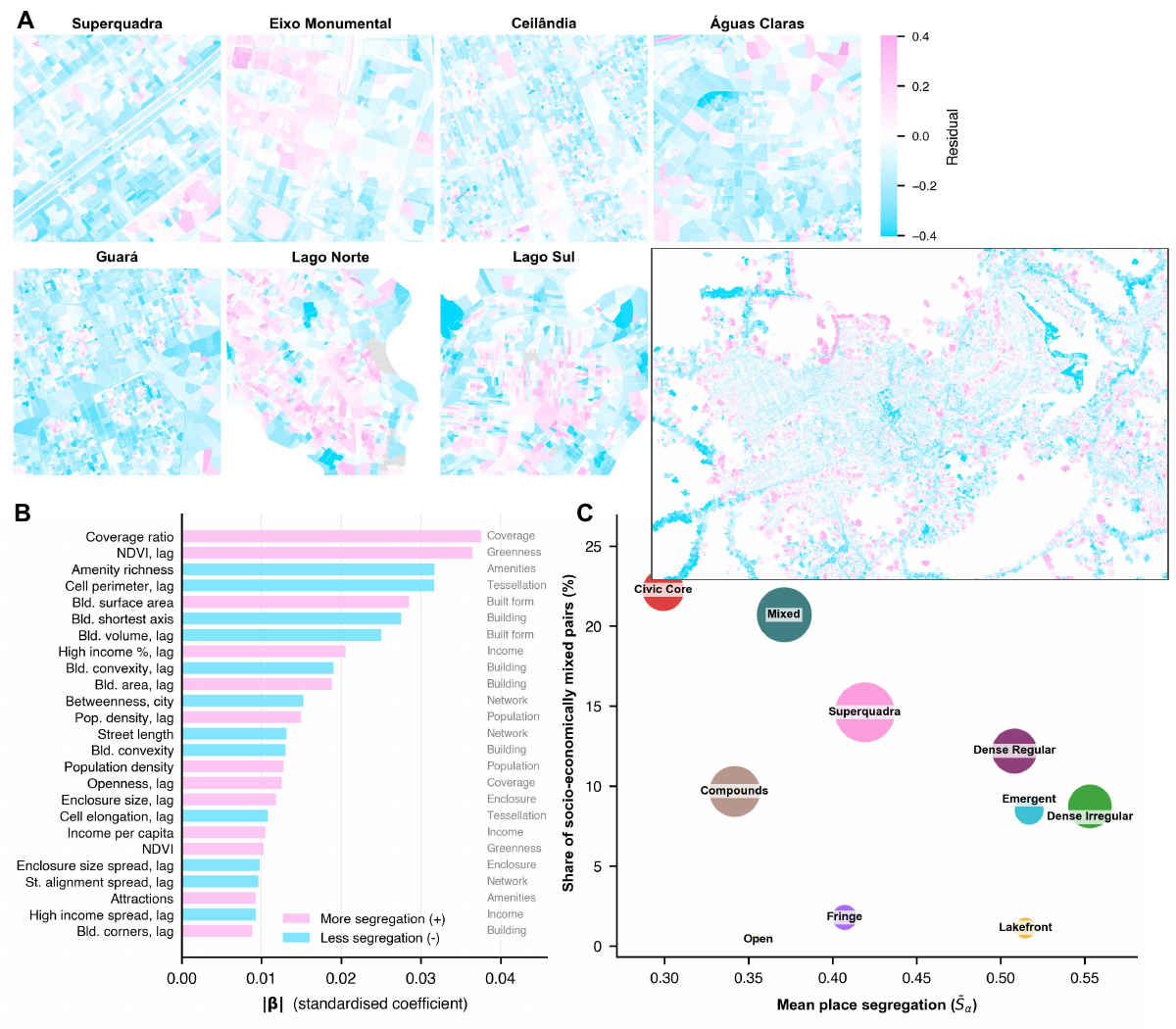}
\caption{\textbf{Modelling the role of form, function and demography.} \textbf{A} Residuals from our model of place segregation $S_{\alpha}$ in selected areas. Positive residuals mark cells that are more segregated than predicted by their local built form and demographic context, while negative residuals mark cells that are more mixed than expected. The lakefront suburbs remain more segregated than the model predicts, whereas several corridor-like and mixed-use areas are less segregated than expected. \textbf{B} Standardised coefficients from the model, with positive coefficients indicating features associated with more segregation, vice versa for negative coefficients. Amenities are associated with lower segregation, while coverage ratio and greener contextual settings are associated with higher segregation, indicating that both dense fabrics and sparse vegetated enclaves can produce separation through different spatial logics. \textbf{C} Comparing the share of all mixed interactions, when a person of one quintile shares the same tessellation at the same time as someone from a different quintile, to place segregation $S_\alpha$, by morphotype: civic cores comprise a plurality of these mixed interactions, but other typologies account for more visits, represented by marker size.}
\label{fig5}
\end{figure*}

The coefficient profiles in Fig. ~\ref{fig5}\textbf{B} point to three broad mechanisms. First, places rich in amenities are generally associated with lower segregation, suggesting that mixed functionality pulls together broader publics. Second, both high coverage ratios and greener, more open contexts are associated with greater segregation, implying two distinct urban paths to separation: dense tissues on the one hand, and sparse, vegetated enclaves on the other. Third, the residual maps in Fig. ~\ref{fig5}\textbf{A} show that these variables capture much, but not all, of Brasília's social geography. The model under-predicts segregation in the lakefront suburbs and over-predicts it in several mixed or corridor-like areas, indicating that adjacency, barrier effects, and metropolitan position continue to matter beyond the measured local features.

We can see that generating interactions, per se, is not enough. Fig. ~\ref{fig5}\textbf{C} shows the relationship between the share of socio-economically mixed interactions, place segregation and the volume of total interactions, per morphotype. Most cross-class co-presences occur in civic cores and mid-rise, mixed-use areas, but dense residential areas still host a large share of all co-presences. Place segregation is higher in dense, poor regular and irregular neighbourhoods that account for 21\% of all interactions. Further, while these dense and poor neighbourhoods host 20\% of all cross-class co-presences, they host just 6.7\% of encounters between top and bottom quintiles. The civic cores host 21.9\% of these rich-poor interactions, despite hosting just 15.8\% of all cross-class co-presences; these values are 22.9\% and 13.5\% for superquadras. 

Taken together, these results show that Brasília's mobility system does not simply dissolve residential segregation into metropolitan mixing. Rather, movement redistributes segregation across a hierarchy of places. The most mixed environments are concentrated in civic and mixed-use nodes, but these are embedded in a larger landscape of affluent enclaves, dense poor fabrics, and residential modernist districts that continue to channel people toward socially familiar encounter fields. Urban form therefore does not determine encounter mechanically, but it does make some forms of mixing more available, and some forms of avoidance much easier to sustain.

\section*{Discussion}
Brasília reveals a specific version of a generic urban fact: segregation occurs not only at home but throughout daily experience. Further, it is produced through the geometry of access. The city's most mixed places are concentrated in civic and commercial nodes, while its residential fabrics---from the affluent lakeside subdivisions to dense poor settlements---channel residents toward more socially familiar encounter fields. This is why the relevant contrast is not simply between mobility and immobility. Most encounters in Brasília occur in these residential zones, so mobility can soften neighbourhood isolation, but it can also reproduce it, especially when the urban system allows some groups to reach desirable destinations while bypassing intervening publics.

These findings also complicate any easy indictment or easy defence of modernist planning. The Plano Piloto's civic core remains relatively permeable, and the superquadras are less segregated than many peripheral residential tissues. But neither outcome licenses the claim that formal openness or planned order are sufficient conditions for social integration. The stronger pattern is functional: places that combine amenities, accessibility, and mixed uses draw broader publics, whereas places organised primarily as residential enclaves remain socially narrower, regardless of whether they are rich or poor, formal or informal. For planned capitals and contemporary new towns, the lesson may be that repeating neighbourhood units or separating functions may reduce certain frictions of movement, but without shared destinations and porous connections they also make selective exposure easier to sustain. Shared destinations in Brasília are critical nodes, but there are few of them: just three of our morphometric classes have any points of interest, and these classes represent just 7\% of all enclosed tessellations---plots with buildings---in the city. 

Our work also suggests that the scale of observation is not a technical afterthought. Brasília can look considerably more integrated when read through coarse districts than when read through the finer street-and-building enclosures that structure everyday access. That is not merely a cartographic curiosity. It means that segregation is fundamentally microscopic: produced where corridors, blocks, barriers, and amenities sort people into distinct fields of encounter. Administrative units can hide that process precisely because they aggregate over it.

Several limitations matter. First, income is inferred from residential location and associated metadata rather than observed directly at the individual level, so our estimates should be read as socio-economic exposure profiles rather than exact interpersonal class contact. Second, the analysis captures one month of GPS traces against a static census backdrop; longer panels would allow cleaner separation of habitual patterns from short-term fluctuations. Third, our segregation measures are based on co-visitation distributions rather than verified social interaction, and our morphotypes are derived from an unsupervised classification whose boundaries are analytically useful rather than ontologically fixed. None of these caveats undo the central result. They do, however, set the scale of the claim: Brasília shows how urban form structures the opportunities for encounter and avoidance, not how every encounter is ultimately experienced or interpreted.

For planners, the lesson is not that order is bad and disorder is good. Brasília contains planned spaces that mix and planned spaces that isolate. The more consequential distinction is between places organised as shared destinations and places organised as residential catchments. Where amenities, employment, services, and public-facing centrality are concentrated in a few nodes, mobility can reduce isolation only unevenly and at unequal cost. Where those destinations are distributed across the metropolitan system and linked through more porous networks, cross-class encounter becomes less exceptional and less burdensome. For Brasília, and for other planned capitals and new towns, the practical agenda is therefore metropolitan rather than merely architectural: reduce barrier effects, strengthen interstitial connectors between nuclei, and grow mixed-use centralities outside the privileged core.


\bibliography{references}

@book{hall2014cities,
  author    = {Hall, Peter},
  title     = {Cities of Tomorrow: An Intellectual History of Urban Planning and Design Since 1880},
  edition   = {4},
  publisher = {Wiley-Blackwell},
  year      = {2014}
}

@book{hall1998sociable,
  author    = {Hall, Peter and Ward, Colin},
  title     = {Sociable Cities: The Legacy of Ebenezer Howard},
  publisher = {John Wiley \& Sons},
  year      = {1998}
}

@book{fishman1982utopias,
  author    = {Fishman, Robert},
  title     = {Urban Utopias in the Twentieth Century: Ebenezer Howard, Frank Lloyd Wright, Le Corbusier},
  publisher = {MIT Press},
  year      = {1982}
}

@book{holston1989modernist,
  author    = {Holston, James},
  title     = {The Modernist City: An Anthropological Critique of Brasília},
  publisher = {University of Chicago Press},
  address   = {Chicago},
  year      = {1989}
}

@book{howard1902garden,
  author    = {Howard, Ebenezer},
  title     = {Garden Cities of To-morrow},
  publisher = {S. Sonnenschein \& Company},
  year      = {1902}
}

@book{peiser2021towns,
  editor    = {Peiser, Richard and Forsyth, Ann},
  title     = {New Towns for the Twenty-First Century: A Guide to Planned Communities Worldwide},
  publisher = {University of Pennsylvania Press},
  year      = {2021}
}

@incollection{perry1929neighborhood,
  author    = {Perry, Clarence A.},
  title     = {The Neighborhood Unit: A Scheme of Arrangement for the Family-Life Community},
  booktitle = {Neighborhood and Community Planning, Regional Plan of New York and Its Environs, Vol. VII},
  publisher = {Committee on Regional Plan of New York and Its Environs},
  year      = {1929},
  pages     = {21--140}
}

@article{silver1985neighborhood,
  author  = {Silver, Christopher},
  title   = {Neighborhood Planning in Historical Perspective},
  journal = {Journal of the American Planning Association},
  year    = {1985},
  volume  = {51},
  number  = {2},
  pages   = {161--174},
  doi     = {10.1080/01944368508976207}
}

@article{lawhon2009determinism,
  author  = {Lawhon, Larry Lloyd},
  title   = {The Neighborhood Unit: Physical Design or Physical Determinism?},
  journal = {Journal of Planning History},
  year    = {2009},
  volume  = {8},
  number  = {2},
  pages   = {111--132},
  doi     = {10.1177/1538513208327072}
}

@article{moreno2021minute,
  author  = {Moreno, Carlos and Allam, Zaheer and Chabaud, Didier and Gall, Catherine and Pratlong, Florent},
  title   = {Introducing the ``15-Minute City'': Sustainability, Resilience and Place Identity in Future Post-Pandemic Cities},
  journal = {Smart Cities},
  year    = {2021},
  volume  = {4},
  number  = {1},
  pages   = {93--111},
  doi     = {10.3390/smartcities4010006}
}

@article{khavariangarmsir2023garden,
  author  = {Khavarian-Garmsir, Amir Reza and Sharifi, Ayyoob and Hajian Hossein Abadi, Mohammad and Moradi, Zahra},
  title   = {From Garden City to 15-Minute City: A Historical Perspective and Critical Assessment},
  journal = {Land},
  year    = {2023},
  volume  = {12},
  number  = {2},
  articleno = {512},
  doi     = {10.3390/land12020512}
}

@article{derntl2019brasilia,
  title     = {Bras{\'\i}lia e seu territ{\'o}rio: a assimila{\c{c}}{\~a}o de princ{\'\i}pios do planejamento ingl{\^e}s aos planos iniciais de cidades-sat{\'e}lites},
  author    = {Derntl, Maria Fernanda},
  journal   = {Cadernos Metr{\'o}pole},
  volume    = {22},
  number    = {47},
  pages     = {123--146},
  year      = {2019}
}

@article{abusaada2022similarity,
  author  = {Abusaada, Hisham and Elshater, Abeer},
  title   = {Examining similarity indicators in six planned capital cities from Africa and Asia: a qualitative research technique},
  journal = {City, Territory and Architecture},
  year    = {2022},
  volume  = {9},
  articleno = {33},
  doi     = {10.1186/s40410-022-00181-2}
}

@article{abusaada2023singularity,
  author  = {Abusaada, Hisham and Elshater, Abeer and Rashed, Rowaida},
  title   = {Exploring the singularity of smart cities in the New Administrative Capital City, Egypt},
  journal = {Ain Shams Engineering Journal},
  year    = {2023},
  volume  = {14},
  number  = {9},
  articleno = {102087},
  doi     = {10.1016/j.asej.2022.102087}
}

@article{syaban2023capital,
  author  = {Syaban, Alfath Satria Negara and Appiah-Opoku, Seth},
  title   = {Building Indonesia's new capital city: an in-depth analysis of prospects and challenges from current capital city of Jakarta to Kalimantan},
  journal = {Urban, Planning and Transport Research},
  year    = {2023},
  volume  = {11},
  number  = {1},
  articleno = {2276415},
  doi     = {10.1080/21650020.2023.2276415}
}

@article{teo2020capital,
  author  = {Teo, Hoong Chen and Lechner, Alex and Sagala, Saut and Campos-Arceiz, Ahimsa},
  title   = {Environmental Impacts of Planned Capitals and Lessons for Indonesia's New Capital},
  journal = {Land},
  year    = {2020},
  volume  = {9},
  number  = {11},
  articleno = {438},
  doi     = {10.3390/land9110438}
}

@misc{reuters2024egypt,
  author       = {{Reuters}},
  title        = {Egypt plans expansion of new capital as first residents trickle in},
  year         = {2024},
  month        = jan,
  day          = {4},
  howpublished = {Reuters}
}

@misc{reuters2026nusantara,
  author       = {{Reuters}},
  title        = {Indonesia's Prabowo affirms commitment to new capital city in first visit as president},
  year         = {2026},
  month        = jan,
  day          = {13},
  howpublished = {Reuters}
}

@article{appleyard1972streets,
  author  = {Appleyard, Donald and Lintell, Mark},
  title   = {The Environmental Quality of City Streets: The Residents' Viewpoint},
  journal = {Journal of the American Institute of Planners},
  year    = {1972},
  volume  = {38},
  number  = {2},
  pages   = {84--101},
  doi     = {10.1080/01944367208977410}
}

@article{saelens2003walking,
  author  = {Saelens, Brian E. and Sallis, James F. and Frank, Lawrence D.},
  title   = {Environmental correlates of walking and cycling: Findings from the transportation, urban design, and planning literatures},
  journal = {Annals of Behavioral Medicine},
  year    = {2003},
  volume  = {25},
  number  = {2},
  pages   = {80--91},
  doi     = {10.1207/S15324796ABM2502_03}
}

@article{evans2003housing,
  author  = {Evans, Gary W. and Wells, Nancy M. and Moch, Annie},
  title   = {Housing and mental health: A review of the evidence and a methodological and conceptual critique},
  journal = {Journal of Social Issues},
  year    = {2003},
  volume  = {59},
  number  = {3},
  pages   = {475--500},
  doi     = {10.1111/1540-4560.00074}
}

@article{guite2006wellbeing,
  author  = {Guite, H. F. and Clark, C. and Ackrill, G.},
  title   = {The impact of the physical and urban environment on mental well-being},
  journal = {Public Health},
  year    = {2006},
  volume  = {120},
  number  = {12},
  pages   = {1117--1126},
  doi     = {10.1016/j.puhe.2006.10.005}
}

@article{frank2004obesity,
  author  = {Frank, Lawrence D. and Andresen, Martin A. and Schmid, Thomas L.},
  title   = {Obesity relationships with community design, physical activity, and time spent in cars},
  journal = {American Journal of Preventive Medicine},
  year    = {2004},
  volume  = {27},
  number  = {2},
  pages   = {87--96},
  doi     = {10.1016/j.amepre.2004.04.011}
}

@article{ewing2003sprawl,
  author  = {Ewing, Reid and Schmid, Tom and Killingsworth, Richard and Zlot, Amy and Raudenbush, Stephen},
  title   = {Relationship between urban sprawl and physical activity, obesity, and morbidity},
  journal = {American Journal of Health Promotion},
  year    = {2003},
  volume  = {18},
  number  = {1},
  pages   = {47--57},
  doi     = {10.4278/0890-1171-18.1.47}
}

@article{liao2025socio,
  title   = {Socio-spatial segregation and human mobility: A review of empirical evidence},
  author  = {Liao, Yuan and Gil, Jorge and Yeh, Sonia and Pereira, Rafael HM and Alessandretti, Laura},
  journal = {Computers, Environment and Urban Systems},
  volume  = {117},
  pages   = {102250},
  year    = {2025}
}

@article{atkinson2016limited,
  title   = {Limited exposure: Social concealment, mobility and engagement with public space by the super-rich in London},
  author  = {Atkinson, Rowland},
  journal = {Environment and Planning A: Economy and Space},
  volume  = {48},
  number  = {7},
  pages   = {1302--1317},
  year    = {2016}
}

@article{davies2019networks,
  title   = {Networks of (dis) connection: mobility practices, tertiary streets, and sectarian divisions in North Belfast},
  author  = {Davies, Gemma and Dixon, John and Tredoux, Colin G and Whyatt, J Duncan and Huck, Jonny J and Sturgeon, Brendan and Hocking, Bree T and Jarman, Neil and Bryan, Dominic},
  journal = {Annals of the American Association of Geographers},
  volume  = {109},
  number  = {6},
  pages   = {1729--1747},
  year    = {2019}
}

@article{rokem2018segregation,
  title   = {Segregation, mobility and encounters in Jerusalem: The role of public transport infrastructure in connecting the ‘divided city’},
  author  = {Rokem, Jonathan and Vaughan, Laura},
  journal = {Urban Studies},
  volume  = {55},
  number  = {15},
  pages   = {3454--3473},
  year    = {2018}
}

@article{schnell2014arab,
  title   = {Arab integration in Jewish-Israeli social space: does commuting make a difference?},
  author  = {Schnell, Izhak and Haj-Yahya, Nasreen},
  journal = {Urban Geography},
  volume  = {35},
  number  = {7},
  pages   = {1084--1104},
  year    = {2014}
}

@article{abdelmonem2015search,
  title   = {In search of common grounds: Stitching the divided landscape of urban parks in Belfast},
  author  = {Abdelmonem, Mohamed Gamal and McWhinney, Rachel},
  journal = {Cities},
  volume  = {44},
  pages   = {40--49},
  year    = {2015}
}

@article{dong2020segregated,
  title   = {Segregated interactions in urban and online space},
  author  = {Dong, Xiaowen and Morales, Alfredo J and Jahani, Eaman and Moro, Esteban and Lepri, Bruno and Bozkaya, Burcin and Sarraute, Carlos and Bar-Yam, Yaneer and Pentland, Alex},
  journal = {EPJ Data Science},
  volume  = {9},
  number  = {1},
  pages   = {20},
  year    = {2020}
}

@article{bokanyi2021universal,
  title   = {Universal patterns of long-distance commuting and social assortativity in cities},
  author  = {Bok{\'a}nyi, Eszter and Juh{\'a}sz, S{\'a}ndor and Karsai, M{\'a}rton and Lengyel, Bal{\'a}zs},
  journal = {Scientific reports},
  volume  = {11},
  number  = {1},
  pages   = {20829},
  year    = {2021}
}

@article{hilman2022socioeconomic,
  title   = {Socioeconomic biases in urban mixing patterns of US metropolitan areas},
  author  = {Hilman, Rafiazka Millanida and I{\~n}iguez, Gerardo and Karsai, M{\'a}rton},
  journal = {EPJ data science},
  volume  = {11},
  number  = {1},
  pages   = {32},
  year    = {2022}
}

@article{moro2021mobility,
  title   = {Mobility patterns are associated with experienced income segregation in large US cities},
  author  = {Moro, Esteban and Calacci, Dan and Dong, Xiaowen and Pentland, Alex},
  journal = {Nature communications},
  volume  = {12},
  number  = {1},
  pages   = {4633},
  year    = {2021}
}

@article{athey2021estimating,
  title   = {Estimating experienced racial segregation in US cities using large-scale GPS data},
  author  = {Athey, Susan and Ferguson, Billy and Gentzkow, Matthew and Schmidt, Tobias},
  journal = {Proceedings of the National Academy of Sciences},
  volume  = {118},
  number  = {46},
  pages   = {e2026160118},
  year    = {2021}
}

@article{browning2022geographic,
  title   = {Geographic isolation, compelled mobility, and everyday exposure to neighborhood racial composition among urban youth},
  author  = {Browning, Christopher R and Tarrence, Jake and Calder, Catherine A and Pinchak, Nicolo P and Boettner, Bethany},
  journal = {American Journal of Sociology},
  volume  = {128},
  number  = {3},
  pages   = {914--961},
  year    = {2022}
}

@article{toth2021inequality,
  title={Inequality is rising where social network segregation interacts with urban topology},
  author    = {T{\'o}th, Gerg{\H{o}} and Wachs, Johannes and Di Clemente, Riccardo and Jakobi, {\'A}kos and S{\'a}gv{\'a}ri, Bence and Kert{\'e}sz, J{\'a}nos and Lengyel, Bal{\'a}zs},
  journal   = {Nature communications},
  volume    = {12},
  number    = {1},
  pages     = {1143},
  year      = {2021}
}

@article{abbiasov202415,
  title     = {The 15-minute city quantified using human mobility data},
  author    = {Abbiasov, Timur and Heine, Cate and Sabouri, Sadegh and Salazar-Miranda, Arianna and Santi, Paolo and Glaeser, Edward and Ratti, Carlo},
  journal   = {Nature Human Behaviour},
  volume    = {8},
  number    = {3},
  pages     = {445--455},
  year      = {2024}
}

@article{aiello2025urban,
  title     = {Urban highways are barriers to social ties},
  author    = {Aiello, Luca Maria and Vybornova, Anastassia and Juh{\'a}sz, S{\'a}ndor and Szell, Michael and Bok{\'a}nyi, Eszter},
  journal   = {Proceedings of the National Academy of Sciences},
  volume    = {122},
  number    = {10},
  pages     = {e2408937122},
  year      = {2025}
}

@article{pinter2025quantifying,
  title    ={Quantifying barriers of urban mobility},
  author   = {Pint{\'e}r, Gerg{\H{o}} and Lengyel, Bal{\'a}zs},
  journal  = {Cities},
  volume   = {167},
  pages    = {106322},
  year     = {2025}
}

@article{derntl2024capitality,
  author  = {Derntl, Maria Fernanda},
  title   = {``Capitality'' beyond the Capital City? Bras{\'i}lia and Its Satellite Towns},
  journal = {Urban History Review},
  volume  = {52},
  number  = {1},
  pages   = {206--235},
  year    = {2024},
  doi     = {10.3138/uhr-2022-0036}
}

@article{cervero1997travel,
  author    = {Cervero, Robert and Kockelman, Kara},
  title     = {Travel demand and the 3Ds: Density, diversity, and design},
  journal   = {Transportation Research Part D: Transport and Environment},
  volume    = {2},
  number    = {3},
  pages     = {199--219},
  doi       = {10.1016/S1361-9209(97)00009-6},
  year      = {1997}
}

@article{ewing2010meta,
  author    = {Ewing, Reid and Cervero, Robert},
  title     = {Travel and the built environment: a meta-analysis},
  journal   = {Journal of the American Planning Association},
  volume    = {76},
  number    = {3},
  pages     = {265--294},
  doi       = {10.1080/01944361003766766},
  year      = {2010}
}

@misc{unesco2026brasilia,
  author       = {{UNESCO World Heritage Centre}},
  title        = {Bras{\'i}lia},
  howpublished = {World Heritage List},
  note         = {Accessed 2026-02-12},
  url          = {https://whc.unesco.org/en/list/445/},
  year         = {2026}
}

@misc{brasil1956novacap,
  author       = {{Brasil}},
  title        = {Lei n. 2.874, de 19 de setembro de 1956 (Cria a Companhia Urbanizadora da Nova Capital do Brasil -- NOVACAP)},
  howpublished = {Planalto, Presid{\^e}ncia da Rep{\'u}blica},
  note         = {Accessed 2026-02-12},
  url          = {https://www.planalto.gov.br/ccivil_03/leis/1950-1969/l2874.htm},
  year         = {1956}
}

@misc{costa1957planopiloto,
  author       = {Costa, L{\'u}cio},
  title        = {Relat{\'o}rio do Plano Piloto de Bras{\'i}lia},
  howpublished = {Relat{\'o}rio apresentado ao concurso nacional para o Plano Piloto da Nova Capital do Brasil},
  note         = {Primary planning document},
  year         = {1957}
}

@misc{costa1987revisitada,
  author       = {Costa, L{\'u}cio},
  title        = {Bras{\'i}lia Revisitada 1985/87: Complementa{\c c}{\~a}o, Preserva{\c c}{\~a}o, Adensamento e Expans{\~a}o Urbana},
  howpublished = {Di{\'a}rio Oficial do Distrito Federal / SINJ-DF},
  note         = {Accessed 2026-02-12},
  url          = {https://www.sinj.df.gov.br/sinj/Diario/1d8c4f0f-a7f4-3bd9-ac37-d01d9a1b7ba3/00fc9851.pdf},
  year         = {1987}
}

@misc{iphan2025superquadra,
  author       = {{Instituto do Patrim{\^o}nio Hist{\'o}rico e Art{\'i}stico Nacional}},
  title        = {A Inven{\c c}{\~a}o da Superquadra},
  howpublished = {Publica{\c c}{\~a}o institucional},
  note         = {Accessed 2026-02-12},
  url          = {https://www.gov.br/iphan/pt-br/superintendencias/distrito-federal/2025-web-ainvencaodasuperquadra_compressed.pdf},
  year         = {2025}
}

@misc{iphan2016patrimonio,
  author       = {{Instituto do Patrim{\^o}nio Hist{\'o}rico e Art{\'i}stico Nacional}},
  title        = {Patrim{\^o}nio em transforma{\c c}{\~a}o: Bras{\'i}lia},
  howpublished = {Publica{\c c}{\~a}o institucional},
  note         = {Accessed 2026-02-12},
  url          = {https://portal.iphan.gov.br/uploads/publicacao/patrimmonio_em_transformacao_braslia_r.pdf},
  year         = {2016}
}

@techreport{ipea2015governanca,
  author      = {{Instituto de Pesquisa Econ{\^o}mica Aplicada}},
  title       = {Governan{\c c}a Metropolitana no Brasil: Distrito Federal},
  institution = {IPEA},
  note        = {Accessed 2026-02-12},
  url         = {https://www.ipea.gov.br/redeipea/images/pdfs/governanca_metropolitana/151103_relatorio_analise_distrito_federal.pdf},
  year        = {2015}
}

@misc{costa2016taguatinga,
  author = {Costa, E. B. da},
  title  = {Taguatinga: a primeira cidade-sat{\'e}lite do Distrito Federal},
  note   = {Biblio 3W (Universitat de Barcelona); accessed 2026-02-12},
  url    = {https://www.ub.edu/geocrit/b3w-1180.pdf},
  year   = {2016}
}

@article{goncalves2024deslocamento,
  author  = {Gon{\c c}alves, C. C. S.},
  title   = {Deslocamento ao trabalho e meios de transporte: uma an{\'a}lise dos grupos populacionais do Distrito Federal},
  journal = {Economia \& Regi{\~a}o},
  note    = {Accessed 2026-02-12},
  url     = {https://ojs.uel.br/revistas/uel/index.php/ecoreg/article/view/48469},
  year    = {2024}
}

@techreport{codeplan2020trabalho,
  author      = {{Companhia de Planejamento do Distrito Federal}},
  title       = {An{\'a}lise espacial do mercado de trabalho do Distrito Federal a partir da PDAD 2018},
  institution = {CODEPLAN},
  note        = {Accessed 2026-02-12},
  url         = {https://www.ipe.df.gov.br/documents/9915964/10220346/NT-An%C3%A1lise-espacial-do-mercado-de-trabalho-do-DF-a-partir-da-PDAD-2018.pdf},
  year        = {2020}
}

@misc{codeplan2019aguasclaras,
  author      = {{Companhia de Planejamento do Distrito Federal}},
  title       = {Pesquisa Distrital por Amostra de Domic{\'i}lios (PDAD) 2018: {\'A}guas Claras},
  institution = {CODEPLAN},
  note        = {Accessed 2026-02-12},
  url         = {https://www.ipe.df.gov.br/documents/9915964/10215637/%25C3%2581guas-Claras.pdf},
  year        = {2019}
}

@misc{distritofederal1992aguasclaras,
  author       = {{Distrito Federal}},
  title        = {Lei n. 385, de 16 de dezembro de 1992 (Autoriza a implanta{\c c}{\~a}o do Bairro {\'A}guas Claras)},
  howpublished = {Sistema Integrado de Normas Jur{\'i}dicas do Distrito Federal},
  note         = {Accessed 2026-02-12},
  url          = {https://www.sinj.df.gov.br/sinj/Diario/92f698cc-0f5c-3d81-93c7-ebf610bca944/57df3d9d.pdf},
  year         = {1992}
}

@article{williams2007brasilia,
  author  = {Williams, Richard J.},
  title   = {Bras{\'i}lia depois de Bras{\'i}lia},
  journal = {Arquitextos},
  number  = {083.00},
  note    = {Vitruvius; accessed 2026-02-12},
  url     = {https://vitruvius.com.br/revistas/read/arquitextos/07.083/251},
  year    = {2007}
}

@book{epstein1973plan,
  author    = {Epstein, David G.},
  title     = {Bras{\'i}lia: Plan and Reality: A Study of Planned and Spontaneous Urban Development},
  publisher = {University of California Press},
  year      = {1973}
}

@article{pereira2021quarteirao,
  author  = {Pereira, Lucas Brasil and Cruz, Luciana Saboia Fonseca},
  title   = {Da cr{\'i}tica \`a superquadra ao quarteir{\~a}o murado: o caso de {\'A}guas Claras em Bras{\'i}lia},
  journal = {P{\'o}s. Revista do Programa de P{\'o}s-Gradua{\c c}{\~a}o em Arquitetura e Urbanismo da FAUUSP},
  volume  = {28},
  number  = {52},
  url     = {https://revistas.usp.br/posfau/en/article/download/175043/173230/497626},
  year    = {2021}
}

@misc{ipedf2025verticalizacao,
  author       = {{Instituto de Pesquisa e Estat{\'i}stica do Distrito Federal}},
  title        = {A verticaliza{\c c}{\~a}o urbana e os limites desej{\'a}veis},
  howpublished = {Artigo institucional},
  note         = {Accessed 2026-02-12},
  url          = {https://www.ipe.df.gov.br/a-verticalizacao-urbana-e-os-limites-desejaveis},
  year         = {2025}
}

@article{manicoba2019regioes,
  author  = {Mani{\c c}oba, R. S.},
  title   = {Cria{\c c}{\~a}o de regi{\~o}es administrativas no Distrito Federal e o planejamento territorial},
  journal = {Tempo - T{\'e}cnica - Territ{\'o}rio},
  volume  = {10},
  number  = {2},
  pages   = {1--30},
  url     = {https://periodicos.unb.br/index.php/ciga/article/download/33529/27132/81390},
  year    = {2019}
}

@misc{codeplan2014guara,
  author      = {{Companhia de Planejamento do Distrito Federal}},
  title       = {Pesquisa Distrital por Amostra de Domic{\'i}lios (PDAD) 2013/2014: Guar{\'a}},
  institution = {CODEPLAN},
  note        = {Accessed 2026-02-12},
  url         = {https://biblioteca.cl.df.gov.br/dspace/bitstream/123456789/1615/12/PDAD_Guar%C3%A1_2013-14.pdf},
  year        = {2014}
}

@misc{ipedf2025guara,
  author      = {{Instituto de Pesquisa e Estat{\'i}stica do Distrito Federal}},
  title       = {Guar{\'a}: Pesquisa Distrital por Amostra de Domic{\'i}lios (PDAD-A) 2024},
  institution = {IPEDF},
  note        = {Accessed 2026-02-12},
  url         = {https://pdad.ipe.df.gov.br/files/reports/guar%C3%A1.pdf},
  year        = {2025}
}

@misc{codeplan2018guara,
  author      = {{Companhia de Planejamento do Distrito Federal}},
  title       = {Estudo Urbano-Ambiental: Regi{\~a}o Administrativa do Guar{\'a}},
  institution = {CODEPLAN},
  note        = {Accessed 2026-02-12},
  url         = {https://www.codeplan.df.gov.br/wp-content/uploads/2018/02/Estudo-Urbano-Ambiental-Guar%C3%A1.pdf},
  year        = {2018}
}

@misc{codeplan2018ceilandia,
  author      = {{Companhia de Planejamento do Distrito Federal}},
  title       = {Estudo Urbano-Ambiental: Regi{\~a}o Administrativa de Ceil{\^a}ndia},
  institution = {CODEPLAN},
  note        = {Accessed 2026-02-12},
  url         = {https://www.codeplan.df.gov.br/wp-content/uploads/2018/02/Estudo-Urbano-Ambiental-Ceil%C3%A2ndia.pdf},
  year        = {2018}
}

@misc{ipedf2025ceilandia,
  author      = {{Instituto de Pesquisa e Estat{\'i}stica do Distrito Federal}},
  title       = {Ceil{\^a}ndia: Pesquisa Distrital por Amostra de Domic{\'i}lios (PDAD-A) 2024},
  institution = {IPEDF},
  note        = {Accessed 2026-02-12},
  url         = {https://pdad.ipe.df.gov.br/files/reports/ceilandia.pdf},
  year        = {2025}
}

@misc{codeplan2018amb,
  author       = {{Companhia de Planejamento do Distrito Federal}},
  title        = {Delimita{\c c}{\~a}o do Espa{\c c}o Metropolitano de Bras{\'i}lia},
  howpublished = {Relat{\'o}rio t{\'e}cnico},
  note         = {Accessed 2026-02-12},
  url          = {https://www.codeplan.df.gov.br/wp-content/uploads/2018/03/Delimita%C3%A7%C3%A3o-do-Espa%C3%A7o-Metropolitano-de-Bras%C3%ADlia-AMB.pdf},
  year         = {2018}
}

@article{aslak2020infostop,
  title       = {Infostop: Scalable stop-location detection in multi-user mobility data},
  author      = {Aslak, Ulf and Alessandretti, Laura},
  journal     = {arXiv preprint arXiv:2003.14370},
  year        = {2020}
}

@article{fleischmann2019momepy,
  title      = {Momepy: Urban morphology measuring toolkit},
  author     = {Fleischmann, Martin},
  journal    = {Journal of Open Source Software},
  volume     = {4},
  number     = {43},
  pages      = {1807},
  year       = {2019}
}

@article{arribas2022spatial,
  title      = {Spatial Signatures-Understanding (urban) spaces through form and function},
  author     = {Arribas-Bel, Daniel and Fleischmann, Martin},
  journal    = {Habitat International},
  volume     = {128},
  pages      = {102641},
  year       = {2022},
  publisher  = {Elsevier}
}

@misc{costa1957relatorio,
  author       = {Costa, Lúcio},
  title        = {Relatório do Plano Piloto de Brasília},
  year         = {1957},
  howpublished = {Plano apresentado ao Concurso Nacional para a Nova Capital do Brasil},
  note         = {Reedição, 4. ed., Brasília: IPHAN-DF, 2018}
}

@book{ferreira2020superquadra,
  author    = {Ferreira, Marcílio Mendes and Gorovitz, Matheus},
  title     = {A invenção da superquadra: o conceito de unidade de vizinhança em Brasília},
  publisher = {IPHAN},
  address   = {Brasília},
  year      = {2020},
  edition   = {2}
}

@misc{unesco2026heritage,
  author       = {{UNESCO World Heritage Centre}},
  title        = {Brasilia},
  year         = {2026},
  howpublished = {World Heritage List entry},
  note         = {Accessed 2026-03-08}
}

@article{costa2019spatial,
  author  = {Costa, Cayo and Lee, Sugie},
  title   = {The Evolution of Urban Spatial Structure in Brasília: Focusing on the Role of Urban Development Policies},
  journal = {Sustainability},
  year    = {2019},
  volume  = {11},
  number  = {2},
  pages   = {553},
  doi     = {10.3390/su11020553}
}

@article{derntl2024satellites,
  author  = {Derntl, Maria Fernanda},
  title   = {Capitality beyond the Capital City? Brasília and Its Satellite Towns},
  journal = {Urban History Review},
  year    = {2024},
  volume  = {52},
  number  = {1},
  pages   = {206--235},
  doi     = {10.3138/uhr-2022-0036}
}

@article{moura2011pioneers,
  author  = {Moura, Cristina Patriota de},
  title   = {Pioneers and Entrepreneurs: Bio/Ethnographic Notes Towards an Anthropology of Urban Growth},
  journal = {Vibrant: Virtual Brazilian Anthropology},
  year    = {2011},
  volume  = {8},
  number  = {2},
  pages   = {502--528},
  doi     = {10.1590/S1809-43412011000200025}
}

@article{peixoto2021casas,
  author  = {Peixoto, Elane Ribeiro and Oliveira, Adriana Mara Vaz de and Waldvogel, Alana Silva},
  title   = {As casas de Ceilândia},
  journal = {Revista Brasileira de Estudos Urbanos e Regionais},
  year    = {2021},
  volume  = {23},
  pages   = {e202104},
  doi     = {10.22296/2317-1529.rbeur.202104pt}
}

@article{pereira2021aguas,
  author  = {Pereira, L. B. and Cruz, L. S. F.},
  title   = {Da cr\'itica \`a superquadra ao quarteir\~ao murado: o caso de {\'A}guas Claras em Bras\'ilia},
  journal = {P\'os. Revista do Programa de P\'os-Gradua\c{c}\~ao em Arquitetura e Urbanismo da FAUUSP},
  year    = {2021},
  volume  = {28}
}

@misc{costa2016satelite,
  author       = {Costa, Everaldo Batista da},
  title        = {Taguatinga: a primeira cidade-satélite do Distrito Federal},
  year         = {2016},
  howpublished = {Biblio 3W: Revista Bibliográfica de Geografía y Ciencias Sociales},
  note         = {Universitat de Barcelona}
}

@mastersthesis{almeida2017zoneamento,
  author  = {Almeida, Nath{\'a}lia Lima de Ara{\'u}jo},
  title   = {Zoneamento: do ideal ao real: externalidades ambientais negativas da ocupação irregular no DF: a experiência do setor habitacional Vicente Pires},
  school  = {Universidade de Brasília},
  address = {Brasília},
  year    = {2017},
  doi     = {10.26512/2017.06.D.24857}
}

@techreport{ipedf2025plano,
  author      = {{Instituto de Pesquisa e Estatística do Distrito Federal}},
  title       = {Pesquisa Distrital por Amostra de Domicílios Ampliada (PDAD-A) 2024: Plano Piloto},
  institution = {IPEDF},
  address     = {Brasília},
  year        = {2025}
}

@techreport{ipedf2025taguatinga,
  author      = {{Instituto de Pesquisa e Estatística do Distrito Federal}},
  title       = {Pesquisa Distrital por Amostra de Domicílios Ampliada (PDAD-A) 2024: Taguatinga},
  institution = {IPEDF},
  address     = {Brasília},
  year        = {2025}
}

@techreport{ipedf2025aguas,
  author      = {{Instituto de Pesquisa e Estatística do Distrito Federal}},
  title       = {Pesquisa Distrital por Amostra de Domicílios Ampliada (PDAD-A) 2024: Águas Claras},
  institution = {IPEDF},
  address     = {Brasília},
  year        = {2025}
}

@techreport{ipedf2025lagosul,
  author      = {{Instituto de Pesquisa e Estatística do Distrito Federal}},
  title       = {Pesquisa Distrital por Amostra de Domicílios Ampliada (PDAD-A) 2024: Lago Sul},
  institution = {IPEDF},
  address     = {Brasília},
  year        = {2025}
}

@techreport{lagosul2020gestao,
  author      = {{Administração Regional do Lago Sul}},
  title       = {Relatório de Gestão da Administração Regional do Lago Sul 2020},
  institution = {Governo do Distrito Federal},
  address     = {Brasília},
  year        = {2020}
}

@techreport{ipedf2025lagonorte,
  author      = {{Instituto de Pesquisa e Estatística do Distrito Federal}},
  title       = {Pesquisa Distrital por Amostra de Domicílios Ampliada (PDAD-A) 2024: Lago Norte},
  institution = {IPEDF},
  address     = {Brasília},
  year        = {2025}
}

@techreport{ipedf2025vicente,
  author      = {{Instituto de Pesquisa e Estatística do Distrito Federal}},
  title       = {Pesquisa Distrital por Amostra de Domicílios Ampliada (PDAD-A) 2024: Vicente Pires},
  institution = {IPEDF},
  address     = {Brasília},
  year        = {2025}
}

@article{kristensen2023urban,
  title={Urban mobility injustice and imagined sociospatial differences in cities},
  author={Kristensen, Nikolaj Grauslund and Lindberg, Malene Rudolf and Freudendal-Pedersen, Malene},
  journal={Cities},
  volume={137},
  pages={104320},
  year={2023},
  publisher={Elsevier}
}

@article{gao2023socio,
  title={Socio-spatial integration in innovation districts: Singapore's mixed-use experiment},
  author={Gao, Tongchaoran and Lim, Samson},
  journal={Cities},
  volume={140},
  pages={104405},
  year={2023},
  publisher={Elsevier}
}

@article{miranda2020shape,
  title={The shape of segregation: The role of urban form in immigrant assimilation},
  author={Miranda, Arianna Salazar},
  journal={Cities},
  volume={106},
  pages={102852},
  year={2020},
  publisher={Elsevier}
}

@article{useche2024spatial,
  title={Spatial segregation patterns and association with built environment features in Colombian cities},
  author={Useche, Andres F and Sarmiento, Olga L and Alvarez-Rivadulla, Maria Jose and Medina, Pablo and Higuera-Mendieta, Diana and Montes, Felipe},
  journal={Cities},
  volume={152},
  pages={105217},
  year={2024},
  publisher={Elsevier}
}

@article{sun2024social,
  title={Social segregation levels vary depending on activity space types: Comparison of segregation in residential, workplace, routine and non-routine activities in Tokyo metropolitan area},
  author={Sun, Chenchen and Shibuya, Yuya and Sekimoto, Yoshihide},
  journal={Cities},
  volume={146},
  pages={104745},
  year={2024},
  publisher={Elsevier}
}

\section*{Acknowledgements}
The author thanks \href{https://locomizer.com/}{Locomizer} for providing the mobility data used in this study. Access to the data was supported by a grant from the Centre for Digital Innovation at University College London.

\section*{Author contributions statement}
\textbf{A.R.} Conceptualisation, methodology, analysis, writing. 

\subsection*{Data and code availability}
We are unable to release the GPS mobility data. All code necessary to reproduce the spatial signatures in Brasília, as well as all figures from aggregated mobility data, will be made available at \href{https://github.com/asrenninger/morphometry}{https://github.com/asrenninger/morphometry}. 

\section*{Competing interests}
The author declares no competing interests.

\beginsupplementaryinformation

\begin{center}
{\LARGE Supplementary Information for\\[0.5\baselineskip]
\textbf{Planning for isolation? The role of urban form and function in shaping mobility in Bras\'ilia}\par}
\vspace{0.9\baselineskip}
{\large Andrew Renninger*\par}
\vspace{0.35\baselineskip}
{{\normalsize $^*$Corresponding author: Andrew Renninger (E-mail: andrew.renninger.12@ucl.ac.uk)}\par}
\end{center}

\clearpage

\section{Histories of Brasília's planned developments}
\label{histories}

\begin{multicols}{2}

\paragraph{Plano Piloto.} Plano Piloto was the original core of Brasília, defined by the 1957 competition-winning plan and inaugurated with the new capital in 1960. It was laid out as the crossing of a monumental east--west axis and a curved residential axis. superquadras and entrequadras were meant to combine housing, education, commerce, and open space within a strict road hierarchy, later described as monumental, residential, gregarious, and bucolic scales. The intended first residents were federal employees and other middle sectors tied to the new capital, while much of the construction and service workforce remained in camps and early satellite settlements because the planned centre did not absorb metropolitan growth. Over time the district became less the whole city than the administrative, symbolic, and heritage core of a much larger urban system. It still concentrates federal functions and a large share of metropolitan employment, even though most of the population lives elsewhere in the Distrito Federal \cite{costa1957relatorio, costa1987revisitada, holston1989modernist, ferreira2020superquadra, unesco2026heritage, costa2019spatial, ipedf2025plano}.

\paragraph{Lago Norte.} Lago Norte took shape on the east side of Lake Paranoá from an adaptation of the original Brasília scheme, as the intended low-rise house sectors were shifted toward the newly formed lakeshore and the peninsula created by the reservoir works. Occupation began between 1960 and 1965 under NOVACAP projects for the northern lakeshore sectors. Early residents included federal employees transferred to the new capital, and a later wave of teachers acquired financed lots in the 1970s, helping consolidate the district. The area then expanded through detached houses, mansions, and a limited activity centre, while later projects such as Taquari and nearby irregular condominiums pushed growth into environmentally sensitive terrain. In the broader Brasília system, Lago Norte became a low-density residential peninsula tied to the Plano Piloto by a small set of access routes and separated from much of the denser western urban fabric \cite{ipedf2025lagonorte, manicoba2019regioes, costa1987revisitada}.

\paragraph{Lago Sul.} Lago Sul emerged on the south margin of Lake Paranoá as part of the controlled occupation of the lake belt around the new capital. Before Brasília the land formed part of antiga Fazenda Gama and associated rural plots. With the capital works it was subdivided and at first linked to housing for NOVACAP personnel, then increasingly to residences for officials and higher-income households attracted by large lots and lake views. Consolidation accelerated in the 1970s as access to the Plano Piloto improved and local commerce and services multiplied, but the district retained a strongly low-density pattern of detached houses, greenery, and curving roads. Because it frames the protected image of Brasília from the south, it remained subject to tighter urban controls than many later expansions. In the metropolitan constellation, Lago Sul is close to the core and central institutions, yet it functions mainly as a residential lakeside enclave rather than as an autonomous subcentre \cite{ipedf2025lagosul, lagosul2020gestao, manicoba2019regioes}.

\paragraph{Taguatinga.} Taguatinga was officially implanted on 5 June 1958, before Brasília's inauguration, to receive population overflow from informal settlements and worker encampments near the construction zone. The toponym predated the new city and referred to local white clay, and the area already had a small occupation along Córrego Cortado. The modern settlement was a rapid state project: lots were opened, first families were settled in the southern portion of the plan, and by 1960 the city already had tens of thousands of residents. It soon accumulated commerce, services, and transport centrality and became the principal western hinge of metropolitan Brasília. Later territorial splits generated or fed Ceilândia, Samambaia, Águas Claras, and Vicente Pires, making Taguatinga one of the major organizing centres of the westward urban field \cite{ipedf2025taguatinga, costa2016satelite, derntl2024satellites, costa2019spatial}.

\paragraph{Guará I and II.} Guará I and II form a continuous district today, but their origins were sequential. Guará I began to be implemented in 1967 as worker housing associated with the residential-industrial sector west of the core, and its early units were produced through collective self-help schemes for NOVACAP employees and workers tied to nearby employment. Guará II followed in 1972 as an expansion with stronger orientation toward stable public employment. The first built fabric was relatively modest, largely horizontal, and closely tied to the labour geography of the new capital. Over time the district changed class profile as many early low-income households sold to better-off newcomers, and later planning revisions intensified apartment construction and verticalization in several sectors. Within metropolitan Brasília, Guará became a hinge settlement between the Plano Piloto and the westward corridor toward Águas Claras and Taguatinga \cite{ipedf2025guara, manicoba2019regioes, moura2011pioneers, costa2019spatial}.

\paragraph{Ceilândia.} Ceilândia was created in 1971 as a planned resettlement city north of Taguatinga under the Campanha de Erradicação de Invasões, whose initials gave the place its name. Its first residents were families transferred from a cluster of informal settlements near the core, and the first large move brought roughly 82,000 people to the new sectors M and N. The initial plan used two crossing axes in a barrel-shaped outline and was soon enlarged with the O, P, Q, and R sectors as provisional occupation hardened into permanent urban fabric. Early housing was often improvised and later consolidated in masonry. Over time the city developed dense local commerce, strong self-building traditions, and an everyday urbanism quite different from the superquadras. In the larger Brasília constellation, Ceilândia became one of the main population centres of the western agglomeration, linked to Taguatinga and Samambaia and tied to the Plano Piloto chiefly through commuting, services, and metropolitan circulation \cite{ipedf2025ceilandia, manicoba2019regioes, peixoto2021casas, derntl2024satellites}.

\paragraph{Águas Claras.} Águas Claras is the main late-twentieth-century infill district between the Plano Piloto and the western cities. Its prehistory lies in planning studies that reserved the area and in the 1980s occupation of former complementary activity lands near Taguatinga, including Areal. The decisive turn came in the early 1990s, when the district was reimagined as a dense middle-income sector able to fill the gap between existing nuclei and generate enough demand for the metro corridor. The enabling law of 1992 and the subsequent plan proposed collective housing, mixed uses, and a more conventional street-and-block urbanity than the superquadra model. The first consolidated residents were in Areal and adjacent occupations; the later wave was predominantly middle-income condominium households. What was built became far more vertical than the early conception, with towers, gated condominiums, commercial strips, and selective active fronts. Within Brasília's broader metropolitan area, Águas Claras now functions as both a residential and commercial subcentre on the westward transport axis \cite{ipedf2025aguas, pereira2021aguas, costa2019spatial}.

\paragraph{Vicente Pires.} Vicente Pires began as a rural production zone of chácaras and small family properties devoted chiefly to horticultural supply for Brasília. In 1989 a formal arrangement with the former Fundação Zoobotânica granted 30-year land-use contracts to hundreds of rural landowners called chacareiros, but the same period also set the stage for rapid urbanization. From the early 1990s, housing demand near the Plano Piloto and high land prices in more established districts drove the informal subdivision and sale of rural plots, and the area filled with horizontal condominiums before infrastructure and legal regularization caught up. The result was an interstitial suburban fabric produced by informal subdivision, incremental building, and later state recognition, rather than a satellite city planned from the outset. Vicente Pires became its own administrative region only in 2009. Within the larger Brasília constellation, it abuts Taguatinga, Águas Claras, and Guará. \cite{ipedf2025vicente, almeida2017zoneamento, costa2019spatial}.

\end{multicols}

\clearpage

\section{Spatial context}
\label{maps}

\begin{figure*}[h!]
\centering
\includegraphics[width=1\textwidth]{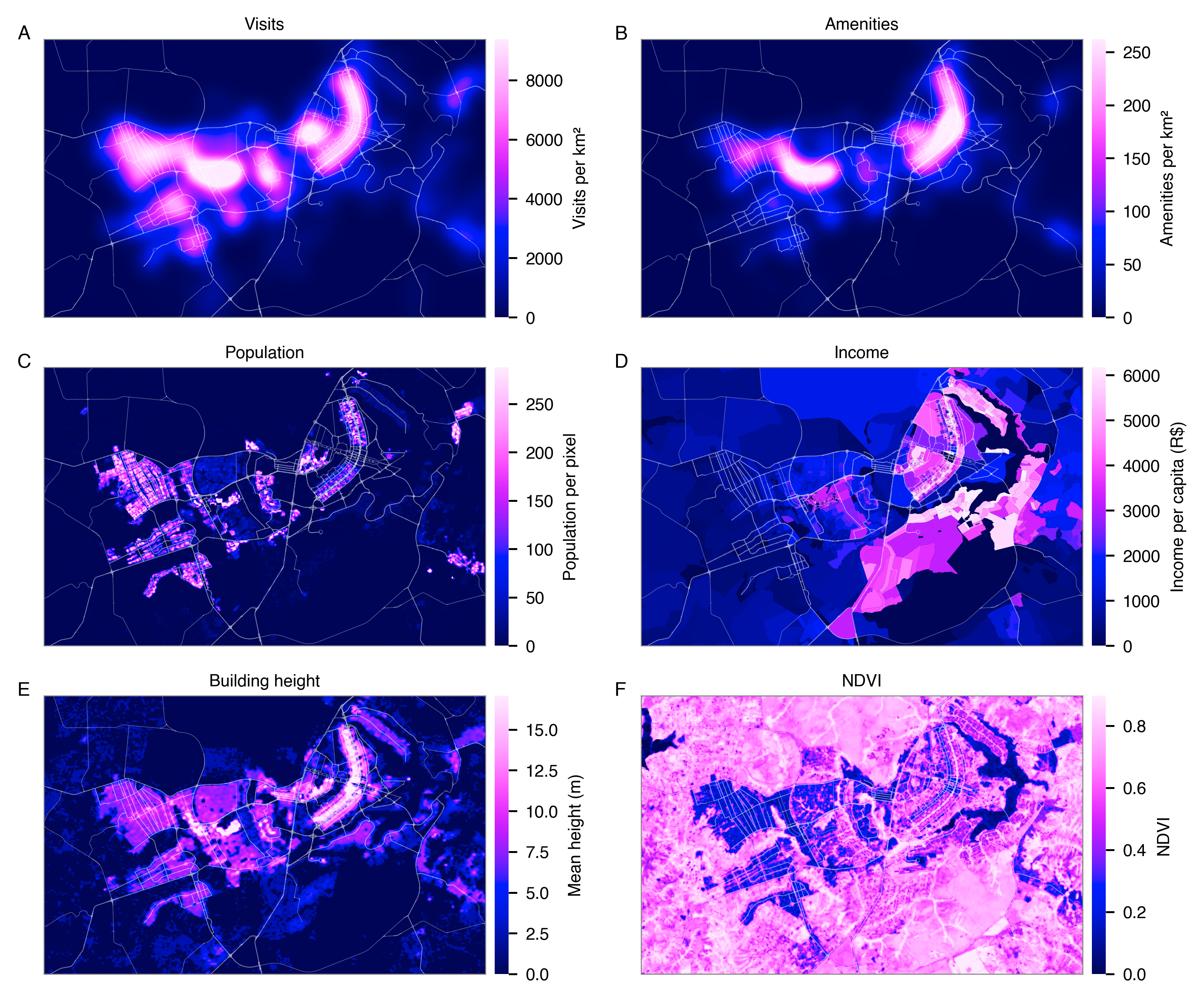}
\caption{\textbf{Amenities, demographics, and structure.} \textbf{A} Kernel density of GPS stays, showing a concentrated but polycentric visitation landscape centred on the Plano Piloto and its satellites. \textbf{B} Amenity density from Overture, likewise concentrated but more so than visits. \textbf{C} Population density, showing that residents are more spatially dispersed and concentrated in peripheral settlements as well as the centre. \textbf{D} Income per capita, with the highest values in the planned core and lakefront sectors. \textbf{E} Mean building height, highest along the central institutional and superblock corridors. \textbf{F} NDVI, highlighting the unusually green planned fabric of the Plano Piloto relative to the denser peripheral settlements. Together these layers show the basic spatial contradiction of Brasília: amenities and visits cluster near the planned core, while population is distributed more broadly across the metropolitan fabric. The amenity and population centres of mass lie just 4.3\,km apart, and place segregation $S_\alpha$ increases with distance from both centres (Pearson $r = 0.283$ and $r = 0.208$, respectively).}
\label{spatial_context}
\end{figure*}

\clearpage

\section{Data validation}
\label{data}

\begin{figure*}[h!]
\centering
\includegraphics[width=1\textwidth]{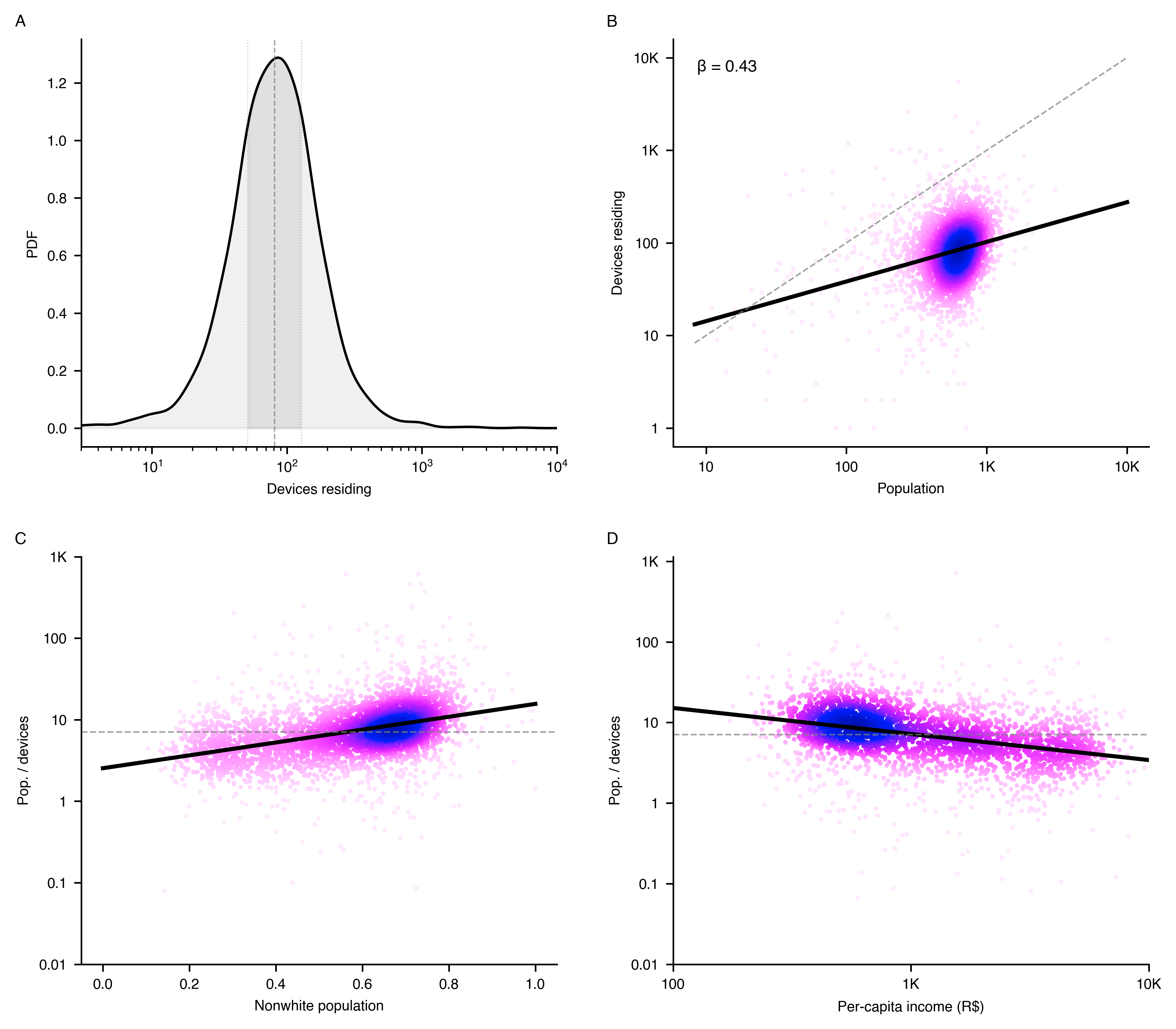}
\caption{\textbf{Coverage and representativeness.} Of 357,901 GPS users, 99.9\% geocode to census tracts, and 72.5\% of populated tracts contain at least one device. \textbf{A} Distribution of devices assigned to tracts, showing a tight concentration around a median of 81 devices per tract (IQR $[51,128]$); because Brazilian tracts are designed to contain roughly 1,000 residents, this corresponds to a median penetration of 141 devices per 1,000 residents. \textbf{B} Relationship between tract population and device count, showing a positive but sublinear association on log--log axes ($\rho = 0.289$, $p < 0.001$), partly because Brazilian tracts have a restricted population range. \textbf{C} Bias check by racial composition: tracts with larger nonwhite shares have slightly fewer devices per capita ($\rho = 0.389$ when expressed as population per device). \textbf{D} Bias check by income: higher-income tracts are somewhat overrepresented ($\rho = -0.349$ for population per device versus income). These biases are modest, and the income bias cuts against our main result---wealthier areas are sampled slightly more heavily, so the observed segregation gradient is conservative rather than inflated.}
\label{data_validation}
\end{figure*}

\clearpage

\section{Morphotypes}
\label{morphotypes}

\begin{figure*}[h!]
\centering
\includegraphics[width=1\textwidth]{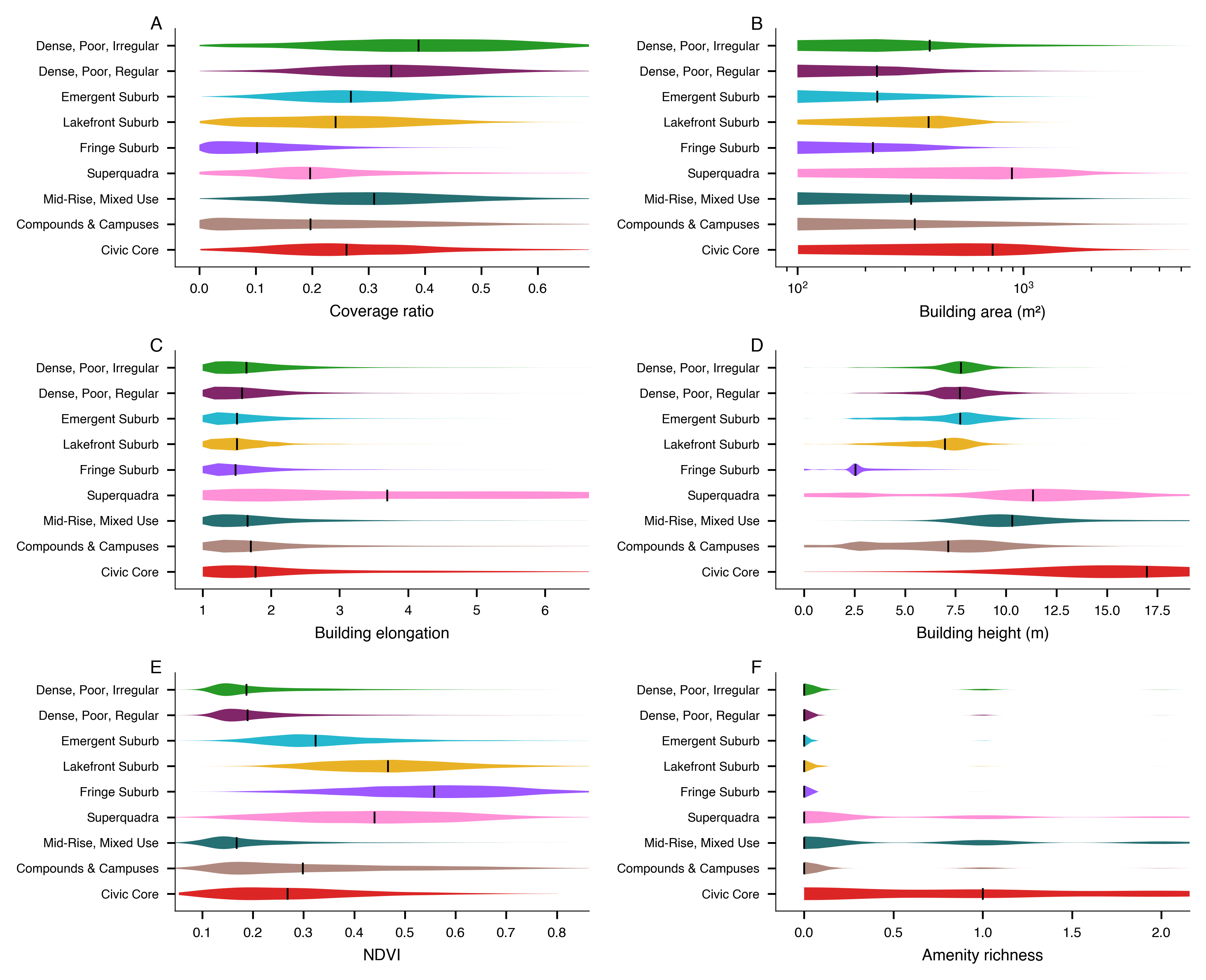}
\caption{\textbf{Morphotypes.} Violin plots of the building-scale signatures that define the nine urban morphotypes used in the analysis (191,095 cells after excluding open, industrial/agricultural, and gap cells). \textbf{A} Coverage ratio. Dense, poor types have the highest coverage, while fringe suburb and superquadra are much more open. \textbf{B} Building area. superquadra and civic core are defined by substantially larger buildings than the peripheral morphotypes. \textbf{C} Building elongation. superquadra stands apart, with large elongated slabs (median $3.69\times$), unlike the compact footprints that dominate elsewhere. \textbf{D} Building height. civic core is tallest (median 17.0\,m), followed by superquadra (11.3\,m) and mid-rise, mixed-use (10.3\,m); fringe suburb is almost entirely low-rise. \textbf{E} NDVI. Planned and lower-density types are greener, especially fringe suburb, lakefront suburb, and superquadra, whereas the dense poor morphotypes have the least vegetation. \textbf{F} Amenity richness. The concentration of amenities features here as well: few morphotypes have any amenities. These distributions show why superquadra is morphologically distinctive: low coverage, very large elongated and tall buildings, surrounded by greenness---the peculiar signature of Brasília's superblocks.}
\label{morphotype_violins_1}
\end{figure*}

\begin{figure*}[h!]
\centering
\includegraphics[width=1\textwidth]{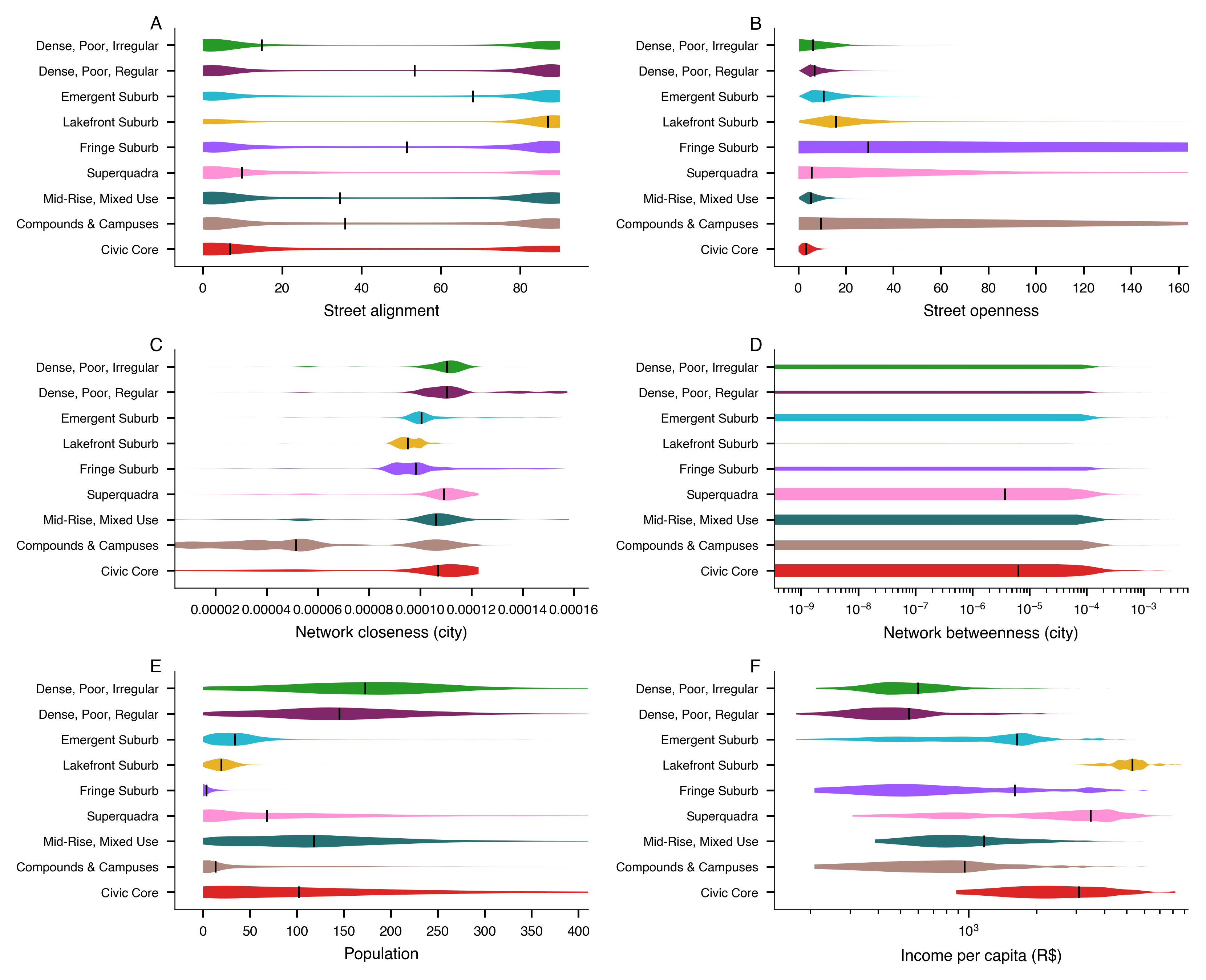}
\caption{\textbf{Morphotypes.} Violin plots of the street-network and socio-demographic signatures of the same nine urban morphotypes. \textbf{A} Street alignment, separating strongly ordered fabrics (especially lakefront suburb, emergent suburb, and dense, poor, regular) from the more jagged layouts of Civic Core, superquadra, and Dense, Poor, Irregular. \textbf{B} Street openness, which is especially right-skewed in Compounds \& Campuses and superquadra because large internal voids coexist with sparse through-streets. \textbf{C} Network closeness, which varies less dramatically than the purely geometric measures but still distinguishes the more peripheral compound-like fabrics from the better-connected central types. \textbf{D} Network betweenness, showing that every morphotype contains a small subset of cells that absorbs disproportionate through-movement. \textbf{E} Population, highest in the dense poor morphotypes and lowest in lakefront suburb and fringe suburb. \textbf{F} Income per capita, with lakefront suburb richest (median R\$5,292), followed by superquadra (R\$3,464) and civic core (R\$3,076), while dense, poor, regular and dense, poor, irregular sit below R\$600. These distributions show that the clusters are social as well as spatial categories: street geometry, network topology, density, and class composition all move together.}
\label{morphotype_violins_2}
\end{figure*}

\clearpage

\section{Sensitivity analyses}

\begin{figure*}[h!]
\centering
\includegraphics[width=1\textwidth]{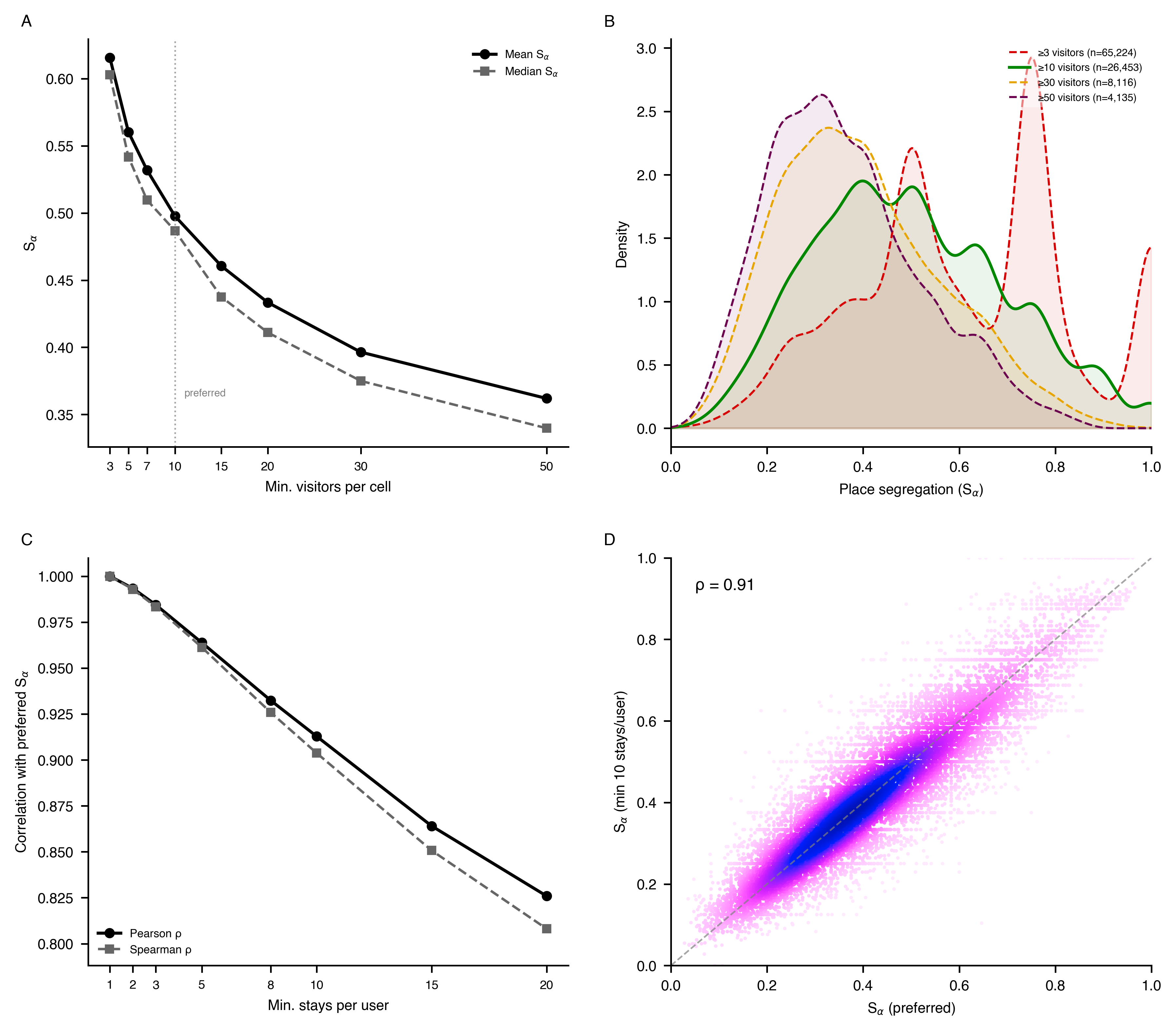}
\caption{\textbf{$S_\alpha$ sensitivity.} \textbf{A} Mean and median place segregation as the minimum number of visitors per cell increases. The preferred specification is 10 visitors per cell (vertical dotted line), which retains 26,453 cells and yields mean $S_\alpha = 0.498$ and median $0.487$. \textbf{B} Density estimates for four cell-traffic thresholds ($\geq 3$, $\geq 10$, $\geq 30$, and $\geq 50$ visitors). For cells shared across thresholds, the values are mathematically identical (all pairwise correlations $= 1.000$); stricter thresholds simply exclude noisier low-traffic cells and shift the reported distribution downward. \textbf{C} Correlation between the preferred specification and increasingly strict minimum-stay filters for users. Filtering users genuinely changes $S_\alpha$, but both Pearson and Spearman correlation remain above 0.80 even at 20 stays per user. \textbf{D} Cell-level comparison between the preferred specification and a stricter filter of at least 10 stays per user ($\rho = 0.91$). The upshot is simple: threshold choice moves levels somewhat, but the geography and rank ordering of segregation survive intact.}
\label{threshold_sensitivity}
\end{figure*}

\begin{figure*}[h!]
\centering
\includegraphics[width=1\textwidth]{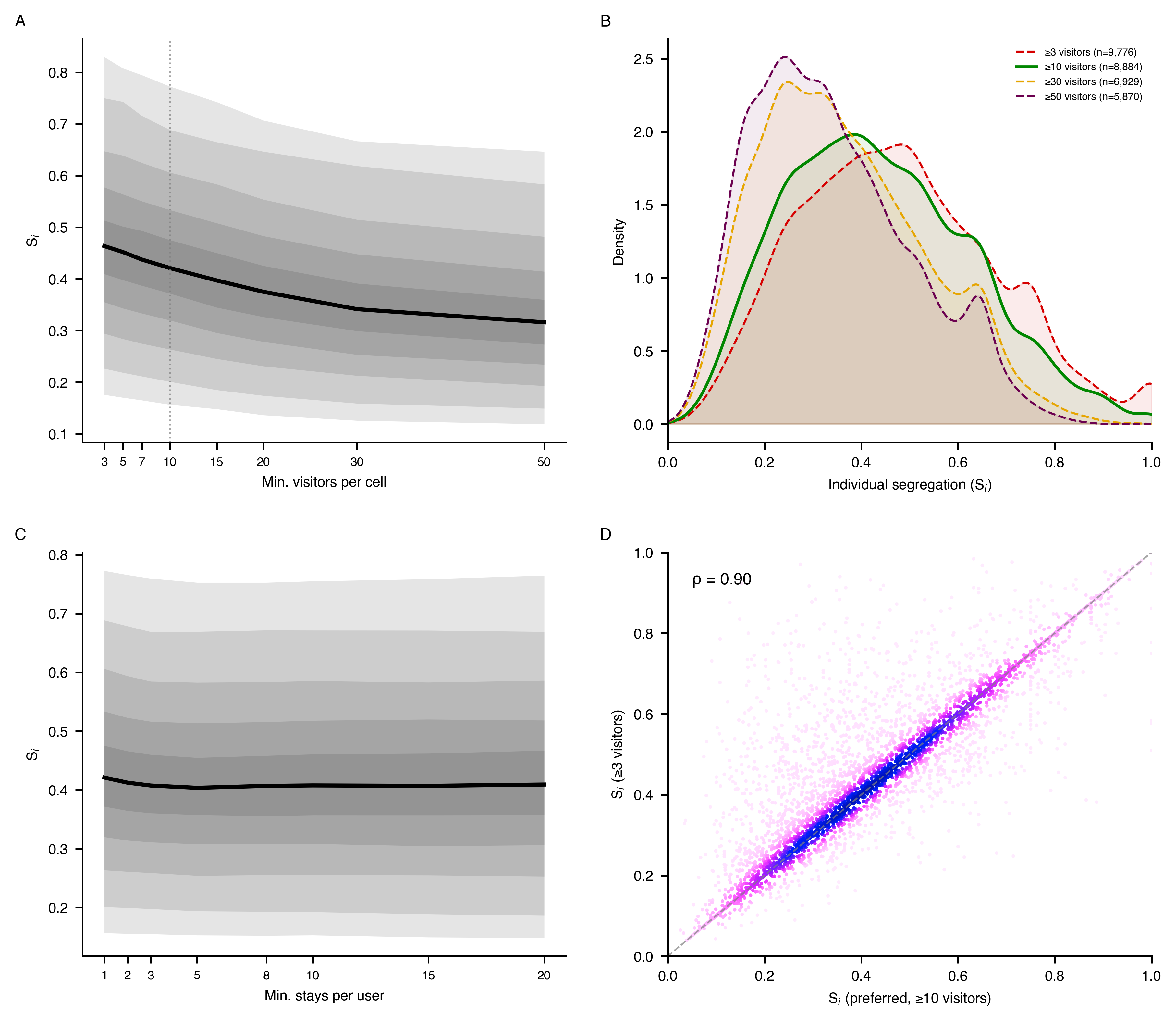}
\caption{\textbf{$S_i$ sensitivity.} \textbf{A} Mean individual segregation as the minimum number of visitors per cell increases, with the preferred threshold of 10 visitors per cell marked by the vertical dotted line. Stricter cell thresholds shift $S_i$ downward because they preferentially retain people whose activity spaces intersect larger, busier, and therefore more mixed locations. \textbf{B} Density estimates of $S_i$ under four cell-traffic thresholds ($\geq 3$, $\geq 10$, $\geq 30$, and $\geq 50$ visitors), showing the same leftward shift. \textbf{C} Mean $S_i$ as the minimum number of stays per user increases. Unlike the cell filter, user-quality filtering barely moves the distribution once a person is observed often enough. \textbf{D} Individual-level comparison between the preferred specification ($\geq 10$ visitors per cell) and a more permissive one ($\geq 3$ visitors), showing strong agreement ($\rho = 0.90$). In other words, the exact level of individual segregation depends somewhat on which cells enter the sample, but the ordering of people is highly stable.}
\label{si_threshold_sensitivity}
\end{figure*}

\clearpage

\section{Changes over the week}

\begin{figure*}[h!]
\centering
\includegraphics[width=1\textwidth]{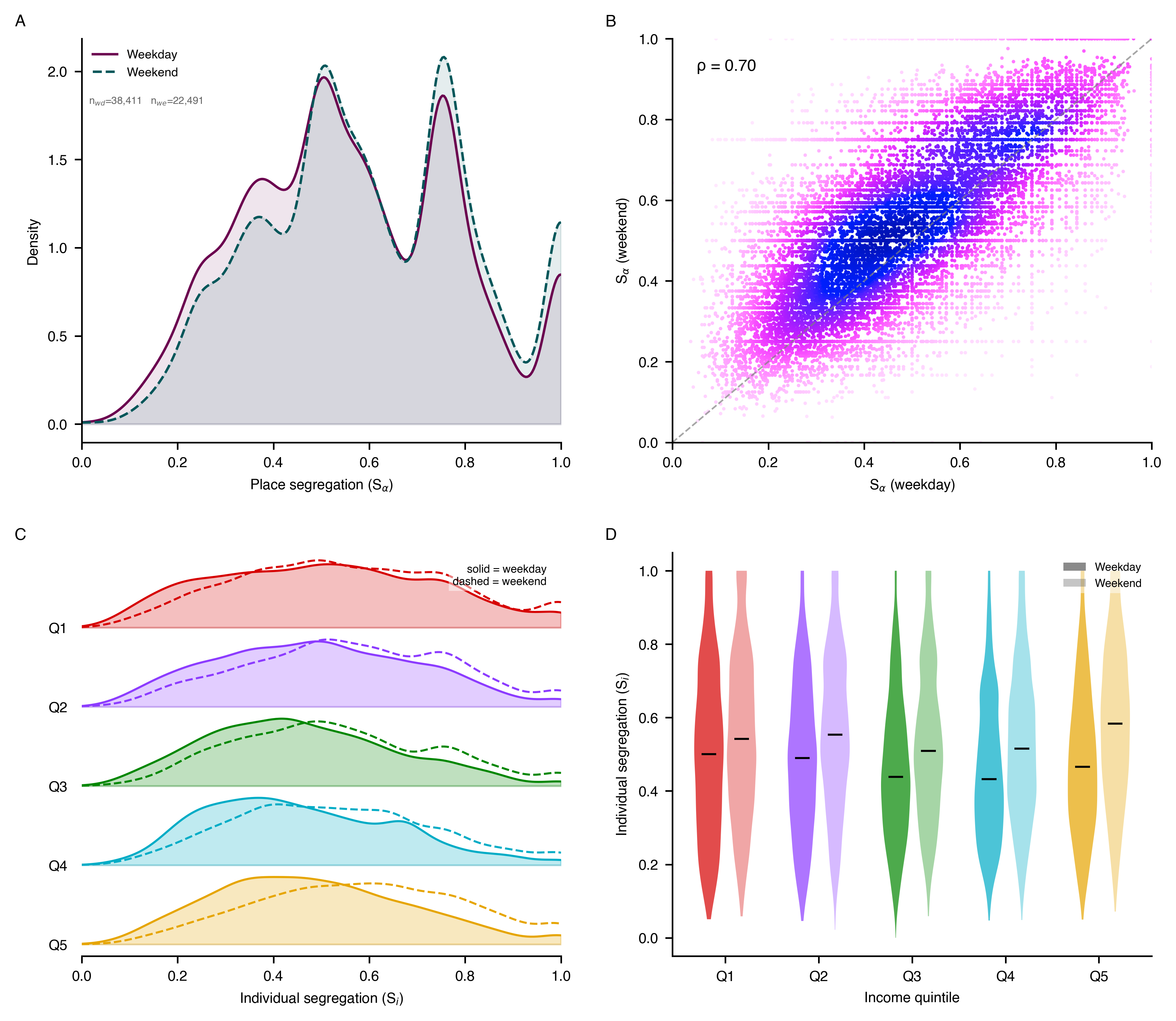}
\caption{\textbf{Weekday--weekend differences.} \textbf{A} Distribution of place segregation $S_\alpha$ on weekdays and weekends. Weekend values are systematically higher (weekday: median $0.542$, mean $0.568$, $n = 38{,}411$; weekend: median $0.583$, mean $0.604$, $n = 22{,}491$; Mann--Whitney $p < 10^{-132}$). \textbf{B} Cell-level comparison of weekday and weekend $S_\alpha$ for the 21,537 cells observed in both periods ($\rho = 0.70$), showing that the weekend shift is broad rather than driven by a few outliers. \textbf{C} Distributions of individual segregation $S_i$ by income quintile, with solid lines for weekdays and dashed lines for weekends. Every quintile shifts right on weekends, implying more homogeneous copresence when commuting ceases. \textbf{D} Equivalent violin plots of $S_i$ by income quintile, showing the same weekend increase and the same U-shaped pattern in which both the poorest and richest users are more individually segregated than the middle quintiles. The interpretation here is straightforward: weekday mobility creates some forced mixing; weekends let residential sorting reassert itself.}
\label{weekday_weekend_segregation}
\end{figure*}
\clearpage

\section{Null model}

\begin{figure*}[h!]
\centering
\includegraphics[width=1\textwidth]{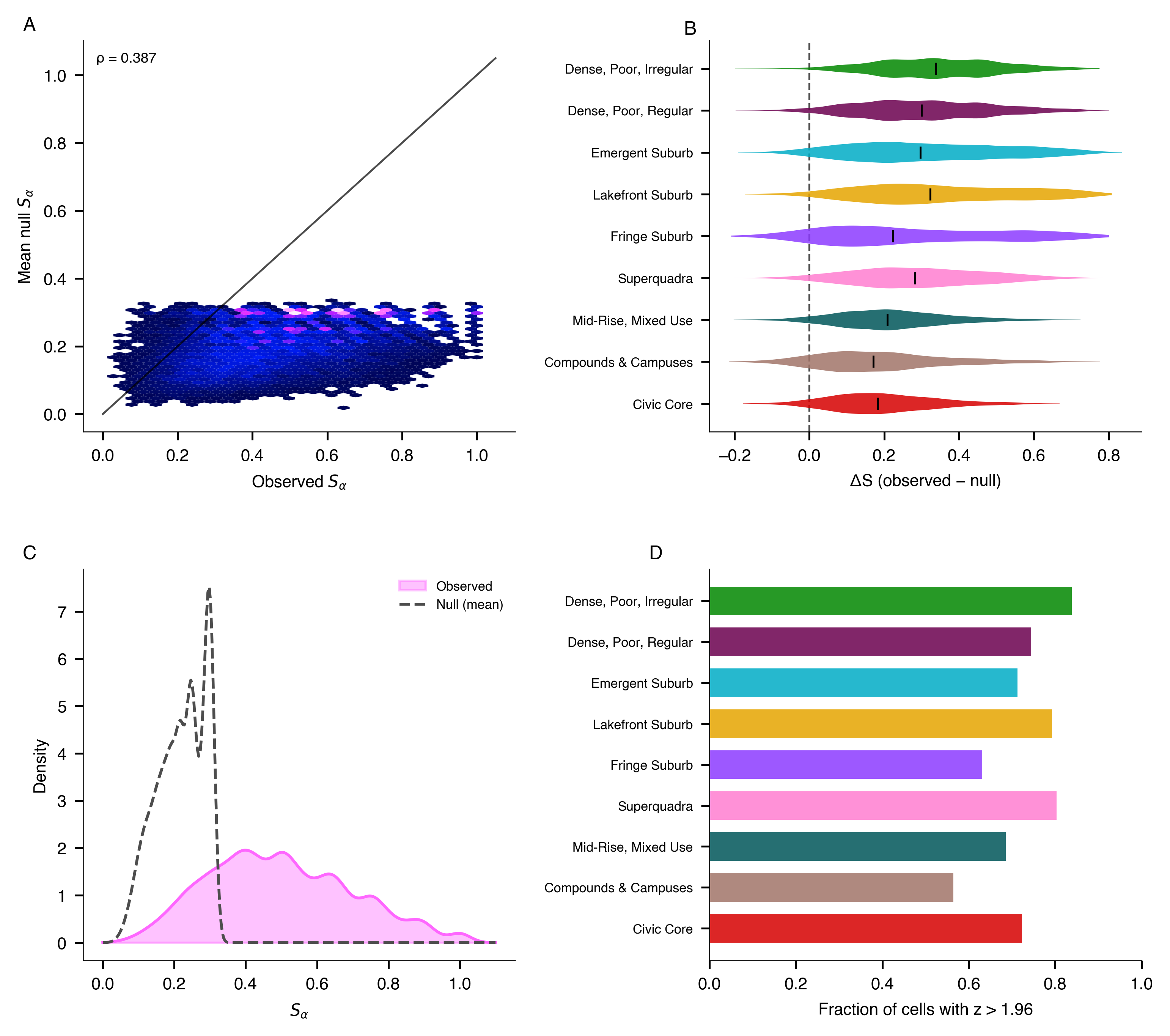}
\caption{\textbf{``Expected'' and observed segregation.} \textbf{A} Cell-level comparison between observed $S_\alpha$ and the null expectation obtained from 1000 permutations that preserve user activity, cell popularity, and distance budget. The weak association ($\rho \approx 0.38$) shows that the null predicts a very different segregation landscape. \textbf{B} Distribution of excess segregation, $\Delta S = S_\alpha^{\mathrm{obs}} - S_\alpha^{\mathrm{null}}$, by morphotype. Every median lies above zero, indicating more segregation than expected under the null across all urban types. \textbf{C} Overall distributions of observed and null $S_\alpha$: the observed landscape is shifted far to the right (observed median $0.487$ versus null median $0.227$; median $\Delta S = 0.260$). \textbf{D} Fraction of cells in each morphotype with $z > 1.96$ relative to the null. Overall, 73.2\% of cells are significantly more segregated than expected and only 0.1\% are significantly less segregated. superquadra exceeds the null in 80.6\% of cells, while Dense, Poor, Irregular reaches 83.8\%. Our null model is simple and by no means captures the manifold pressures that drive mobility, but it shows that segregation is not a trivial byproduct of visit volumes or activity budgets.}
\label{null_model_segregation}
\end{figure*}

\clearpage

\section{Who moves to whom?}

\begin{figure*}[h!]
\centering
\includegraphics[width=1\textwidth]{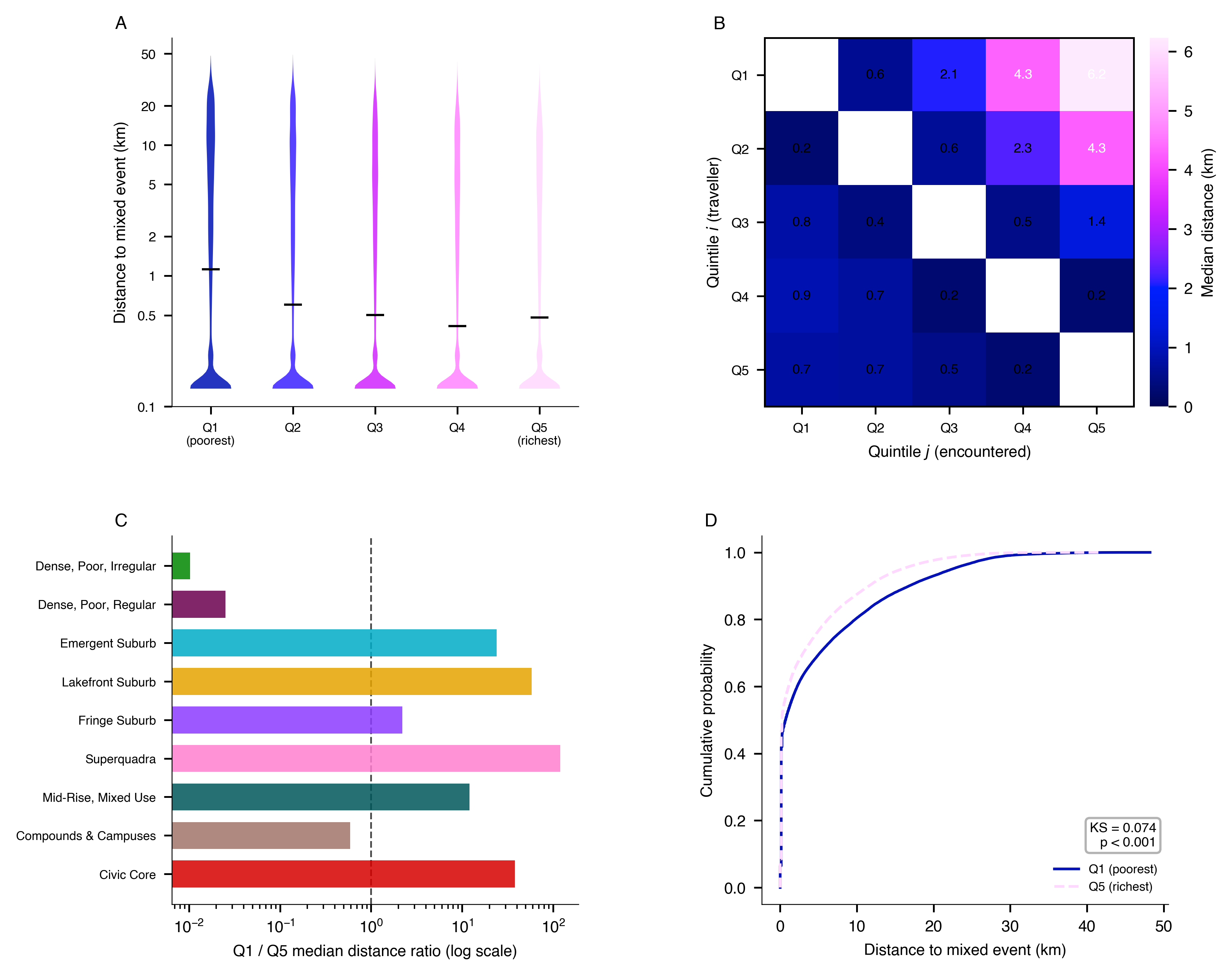}
\caption{\textbf{The burden of mixing.} \textbf{A} Distance travelled to a mixed-class co-presence event by income. The poorest quintile travels farthest (median $0.70\,\mathrm{km}$) whereas every other quintile has a median of $0.24\,\mathrm{km}$; Q1 therefore travels $2.87\times$ farther than Q5 to achieve cross-class contact. \textbf{B} Median travel distance from traveller quintile $i$ to encountered quintile $j$. Poor-to-rich mixing is the most spatially burdensome: Q1 travellers move 4.3\,km to encounter Q4 and 6.2\,km to encounter Q5, whereas Q5 travellers move only 0.7\,km to encounter Q1 and 0.2\,km to encounter Q4. \textbf{C} Q1/Q5 median distance ratio by morphotype; the dashed line marks parity. The asymmetry is extreme in planned areas---$119\times$ in superquadra, $58\times$ in lakefront suburb, and $38\times$ in Civic Core---but reverses in poor morphologies, where rich visitors must travel into predominantly low-income areas. \textbf{D} Cumulative distributions of distance to mixed-class events for Q1 and Q5 (KS $= 0.074$, $p < 0.001$). Brasília does not mix symmetrically: the poor do most of the moving.}
\label{burden_of_mixing}
\end{figure*}

\clearpage

\section{Barriers to mobility}

\begin{figure*}[h!]
\centering
\includegraphics[width=1\textwidth]{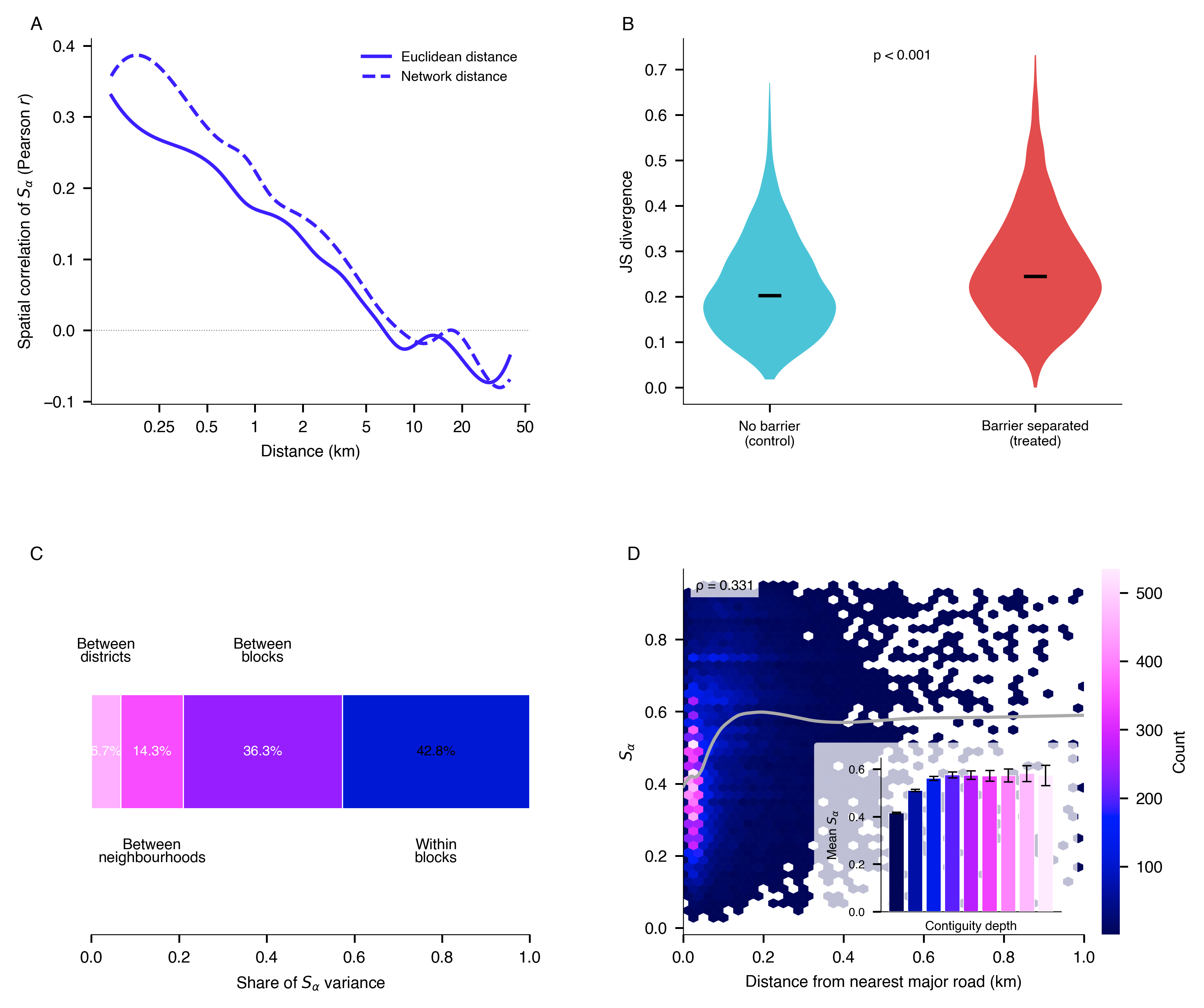}
\caption{\textbf{The role of physical barriers in segregation.} \textbf{A} Spatial correlogram of place segregation $S_\alpha$ using Euclidean distance and travel distance along the road network. The network-based correlogram is uniformly higher, showing that cells close in network space are more socially similar than cells equally close in straight-line space; the median network detour ratio is 1.24. \textbf{B} Jensen--Shannon divergence in income composition between distance-matched cell pairs separated by a barrier (treated) or not (control). Barrier-separated pairs are more dissimilar (median 0.245 versus 0.203; Mann--Whitney $p < 0.001$), a 20.5\% increase at the same Euclidean distance. \textbf{C} Hierarchical variance decomposition of $S_\alpha$ across the enclosure hierarchy. Districts explain 6.7\% of variance, neighbourhoods add 14.3\%, blocks add 36.3\%, and 42.8\% remains within blocks; in total, 57.2\% of segregation variance sits at or above the block level. \textbf{D} $S_\alpha$ rises with distance from the nearest primary, secondary, or tertiary road ($\rho = 0.331$). The inset shows mean $S_\alpha$ by contiguity depth from the road network, with a sharp jump from 0.415 in road-fronting cells to 0.510 one cell inward, then a plateau around 0.57. Roads, in other words, behave as mixing edges; the interiors they enclose behave as segregated cores.}
\label{barriers_matter}
\end{figure*}

\clearpage

\section{Model reporting}
\label{models}

\begin{table*}[h!]
\centering
\caption{\textbf{Elastic-net model summary.} Model fit, sparsity, and regularisation regime for the visitation and segregation specifications. The main contrast is structural: the visits model is near-Ridge, while the segregation models are near-LASSO.}
\label{tab:enet_summary}
\begin{threeparttable}
\footnotesize
\setlength{\tabcolsep}{4.5pt}
\begin{tabularx}{\textwidth}{@{}llYcccccc@{}}
\toprule
\multicolumn{3}{c}{Model} & \multicolumn{6}{c}{Fit and sparsity} \\
\cmidrule(lr){1-3}\cmidrule(lr){4-9}
Family & ID & Specification & Cells & Predictors & $R^2$ & $l_1$ ratio & Non-zero & Active share \\
\midrule
Visitation & DV1 & $\log(\mathrm{visits}) \sim$ urban features & 26,453 & 121 & 0.436 & 0.10 & 82 & 67.8\% \\
\midrule
\multirow{3}{*}{Segregation}
& DV2  & $S_{\alpha} \sim$ urban features & 26,453 & 121 & 0.363 & 0.99 & 61 & 50.4\% \\
& DV3  & $S_{\alpha} \sim$ urban features $+ \log(\mathrm{visits})$ & 26,453 & 122 & 0.378 & 0.99 & 66 & 54.1\% \\
& DV3b & $S_{\alpha} \sim \log(\mathrm{visits})$ only & 26,453 & 1 & 0.149 & --- & 1 & 100.0\% \\
\bottomrule
\end{tabularx}
\begin{tablenotes}[flushleft]
\footnotesize
\item Notes: All 121 urban predictors were standardised before estimation. Models were fit with \texttt{ElasticNetCV}, using 5-fold cross-validation to select both the penalty strength and the LASSO--Ridge mixture ($l_1$ ratio). Lower $l_1$ ratios are more Ridge-like; higher values are more LASSO-like.
\end{tablenotes}
\end{threeparttable}
\end{table*}

\clearpage

\begin{table*}[h!]
\centering
\caption{\textbf{Leading standardised coefficients in the visits and segregation models.} Variables are reported using the exact feature names from the modelling pipeline. Shared predictors appearing in both top-10 lists are marked with $\dagger$.}
\label{tab:enet_top_coefs}
\begin{threeparttable}
\scriptsize
\setlength{\tabcolsep}{4pt}
\begin{tabularx}{\textwidth}{@{}cYcY@{}}
\toprule
Rank & Feature & $\beta$ & Substantive reading \\
\midrule
\multicolumn{4}{l}{\textit{Panel A. DV1: }\(\log(\mathrm{visits})\) \textit{ on urban features}} \\
\midrule
1 & \texttt{bld\_longest\_axis} & +0.136 & Larger buildings attract more visits. \\
2 & \texttt{amenities\_richness}$^{\dagger}$ & +0.127 & More amenities attract more visits. \\
3 & \texttt{bld\_shortest\_axis}$^{\dagger}$ & +0.072 & Wider buildings attract more visits. \\
4 & \texttt{street\_length\_total} & +0.066 & More street network is associated with more visits. \\
5 & \texttt{ghsl\_height\_ctx\_iqm} & +0.062 & Taller surrounding context attracts more visits. \\
6 & \texttt{ndvi\_ctx\_iqm}$^{\dagger}$ & $-0.059$ & Greener surrounding context attracts fewer visits. \\
7 & \texttt{coverage\_ratio}$^{\dagger}$ & $-0.052$ & Denser built coverage attracts fewer visits. \\
8 & \texttt{ghsl\_volume} & $-0.052$ & Larger building volume is associated with fewer visits. \\
9 & \texttt{bld\_convexity} & +0.052 & More convex buildings attract more visits. \\
10 & \texttt{open\_space\_ratio} & +0.052 & More open space is associated with more visits. \\
\addlinespace[3pt]
\midrule
\multicolumn{4}{l}{\textit{Panel B. DV2: }\(S_{\alpha}\) \textit{ on urban features}} \\
\midrule
1 & \texttt{coverage\_ratio}$^{\dagger}$ & +0.038 & Denser cells are more segregated. \\
2 & \texttt{ndvi\_ctx\_iqm}$^{\dagger}$ & +0.036 & Greener surrounding context is more segregated. \\
3 & \texttt{amenities\_richness}$^{\dagger}$ & $-0.032$ & More amenities are associated with less segregation. \\
4 & \texttt{cell\_perimeter\_ctx\_iqm} & $-0.032$ & Larger surrounding cells are associated with less segregation. \\
5 & \texttt{ghsl\_surface} & +0.029 & More built surface is associated with more segregation. \\
6 & \texttt{bld\_shortest\_axis}$^{\dagger}$ & $-0.028$ & Wider buildings are associated with less segregation. \\
7 & \texttt{ghsl\_volume\_ctx\_iqm} & $-0.025$ & Larger surrounding building volume is associated with less segregation. \\
8 & \texttt{income\_high\_pct\_ctx\_iqm} & +0.021 & High-income surrounding context is associated with more segregation. \\
9 & \texttt{bld\_convexity\_ctx\_iqm} & $-0.019$ & More convex surrounding buildings are associated with less segregation. \\
10 & \texttt{bld\_area\_ctx\_iqm} & +0.019 & Larger surrounding building footprints are associated with more segregation. \\
\bottomrule
\end{tabularx}
\begin{tablenotes}[flushleft]
\footnotesize
\item Notes: Coefficients are standardised elastic-net coefficients. Positive signs indicate higher predicted visits or higher predicted segregation, respectively. The visits model is diffuse in magnitude (top coefficient 0.136; 10th coefficient 0.052), whereas the segregation model is flatter and sparser (top coefficient 0.038; 10th coefficient 0.019).
\end{tablenotes}
\end{threeparttable}
\end{table*}

\clearpage

\begin{table*}[h!]
\centering
\caption{\textbf{Cross-model contrasts and mediation diagnostics.} Shared predictors often flip sign between visitation and segregation, and adding visit volume to the morphology model yields only a small gain in explanatory power.}
\label{tab:enet_contrasts}
\begin{threeparttable}
\footnotesize
\setlength{\tabcolsep}{4pt}
\begin{tabularx}{\textwidth}{@{}lcccY@{}}
\toprule
\multicolumn{5}{l}{\textit{Panel A. Model-level contrasts}} \\
\midrule
Metric & DV1 & DV2 & DV3 & Interpretation \\
\midrule
Outcome & $\log(\mathrm{visits})$ & $S_{\alpha}$ & $S_{\alpha}$ & DV1 predicts footfall; DV2 and DV3 predict place segregation. \\
$l_1$ ratio & 0.10 & 0.99 & 0.99 & Visits are near-Ridge; segregation remains near-LASSO even when visits are added. \\
Active coefficients & 82 / 121 & 61 / 121 & 66 / 122 & Segregation is materially sparser than visitation. \\
$R^2$ & 0.436 & 0.363 & 0.378 & Built form already explains most of the modelled variation in segregation. \\
Largest reported $|\beta|$ & 0.136 & 0.038 & 0.033 on $\log(\mathrm{visits})$ & Segregation is not dominated by a single giant exposure term. \\
\addlinespace[3pt]
\midrule
\multicolumn{5}{l}{\textit{Panel B. Mediation / exposure decomposition}} \\
\midrule
Comparison & \multicolumn{3}{c}{Change in fit} & Interpretation \\
\midrule
DV2 $\rightarrow$ DV3 & \multicolumn{3}{c}{$0.363 \rightarrow 0.378$ \; (\(\Delta R^2 = +0.015\))} & Adding visit volume to the full morphology model yields only a small improvement. \\
DV3b $\rightarrow$ DV2 & \multicolumn{3}{c}{$0.149 \rightarrow 0.363$ \; (\(\Delta R^2 = +0.213\))} & Replacing visits-only exposure with morphology more than doubles explained variance ($2.4\times$). \\
DV3b $\rightarrow$ DV3 & \multicolumn{3}{c}{$0.149 \rightarrow 0.378$ \; (\(\Delta R^2 = +0.229\))} & The full model's explanatory power is overwhelmingly carried by built form. \\
\addlinespace[3pt]
\midrule
\multicolumn{5}{l}{\textit{Panel C. Selected shared predictors with opposite signs in DV1 and DV2}} \\
\midrule
Predictor & DV1 $\beta$ & DV2 $\beta$ & Directional contrast & Interpretation \\
\midrule
\texttt{amenities\_richness} & +0.127 & $-0.032$ & attracts visits / reduces segregation & Mixed-use destinations pull people in and make cells more socially mixed. \\
\texttt{coverage\_ratio} & $-0.052$ & +0.038 & reduces visits / increases segregation & Denser coverage is associated with quieter but more homogeneous places. \\
\texttt{ndvi\_ctx\_iqm} & $-0.059$ & +0.036 & reduces visits / increases segregation & Greener context carries the superblock signal: pleasant, but socially sorted. \\
\texttt{bld\_shortest\_axis} & +0.072 & $-0.028$ & attracts visits / reduces segregation & Wider buildings are associated with busier, less segregated cells. \\
\bottomrule
\end{tabularx}
\begin{tablenotes}[flushleft]
\footnotesize
\item Notes: The updated SI report gives full top-10 coefficient rankings for DV1 and DV2, but only the fit summary and the coefficient on $\log(\mathrm{visits})$ for DV3. This table reports only the quantities explicitly documented there.
\end{tablenotes}
\end{threeparttable}
\end{table*}

\clearpage

\end{document}